\documentclass[prc,aps,floats,floatfix,showpacs,nofootinbib,superscriptaddress]{revtex4}

\usepackage{amsmath}
\usepackage{amssymb}
\usepackage{amsfonts}
\usepackage{graphicx}
\usepackage{tabularx}
\usepackage{multirow}
\usepackage{lscape}
\usepackage{rotating}
\usepackage{pstricks}
\usepackage{feynmf}
\usepackage{feynmf}
\usepackage{ dsfont }

\begin{document}

\title{Linear response theory in asymmetric nuclear matter for Skyrme functionals including spin-orbit and tensor terms II: Charge Exchange}


\author{D. Davesne}
\email{davesne@ipnl.in2p3.fr}
\affiliation{Universit\'e de Lyon, F-69003 Lyon, France; Universit\'e Lyon 1,
             43 Bd. du 11 Novembre 1918, F-69622 Villeurbanne cedex, France\\
             CNRS-IN2P3, UMR 5822, Institut de Physique des deux Infinis de Lyon}
\author{A. Pastore}
\email{alessandro.pastore@york.ac.uk}
\affiliation{Department of Physics, University of York, York Y010 5DD, UK}
\author{J. Navarro}
\email{navarro@ific.uv.es}
\affiliation{IFIC (CSIC-Universidad de Valencia), Apartado Postal 22085, E-46.071-Valencia, Spain}


\begin{abstract}
We present the formalism of linear response theory both at zero and finite temperature in the case of asymmetric nuclear matter excited by an isospin flip probe. 
The particle-hole interaction is derived from a general Skyrme functional that includes spin-orbit and tensor terms. Response functions are obtained by solving a closed algebraic system of equations. Spin strength functions are analysed for typical values of density, momentum transfer, asymmetry and temperature. 
We evaluate the role of statistical errors related to the uncertainties of the coupling constants of the Skyrme functional and thus determine the confidence interval of the resulting response function.
\\
\end{abstract}


 
\date{\today}


\maketitle


\section{Introduction}

Transport properties of neutrinos play a crucial role in understanding the realisation of several astrophysical scenarios as supernovae-explosions, neutron star mergers or the evolution of protoneutron stars~\cite{hir87,mez05,jan12,bur13,bur86,may87,kei94}. The neutrino mean free path (NMFP) within dense nuclear matter at finite temperature is thus a key ingredient to understand the behaviour of these astrophysical objects.
When the neutrino pass trough the various layers of nuclear matter several processes may take place as elastic scattering or absorption.  In their seminal article, Iwamoto and Pethick~\cite{iwa82} showed that the NMFP could change by a factor of 2 to 3 in a range of densities around saturation since the neutrino can excite a collective nuclear mode and thus lose energy and momentum substantially.

In  a previous series of articles~\cite{pas12a,pas14bsk,report}, we have studied the properties of NMFP using Skyrme functionals~\cite{per04} for pure neutron matter (PNM) at both zero and finite temperature. However, the PNM assumption for a stellar medium is not realistic since a non-negligible fraction of protons is usually present~\cite{lattimer2000nuclear,steiner2005isospin,lattimer2007neutron} as well. Considering an environment including both protons and neutrons is thus crucial.

In Ref. ~\cite{red98}, the authors have studied both neutron and charged reaction rates of neutrinos in dense nuclear matter. The work included only partially the \emph{in-medium} effect at single particle level only, thus neglecting possible collective modes. The main outcome of the article is that the NMFP is dominated by the reaction $\nu_{e^-}+n\rightarrow p+e^-$. Such a  result is also confirmed by other authors using different types of approximations~\cite{she03,cow04,rob12}.
Since NMFP is used for example in neutrino transport radiation of hydrodynamics simulations  ~\cite{pin12}, the exact value of such a quantity has a direct impact on other relevant astrophysical observables as neutrino luminosities.

A fully quantitative calculation of the NMFP requires the knowledge of the nuclear strength function, which is usually determined via the random phase approximation (RPA) or linear response (LR) formalism, based on the particle-hole (ph) interaction between particles below and above the Fermi level. 
The LR of asymmetric nuclear matter to isospin-flip probes has been calculated in Ref.~\cite{her99} using a zero-range Skyrme interaction restricted to its central part, including both direct and exchange terms. We present here results based on a general Skyrme interaction, which also includes  spin-orbit and tensor components. This is of fundamental importance since the latter has been shown to play a significant role in affecting the nuclear response function~\cite{dav09,pas12,pas12a,pas12b,report,pas13b,dav14b}. 

The article is organised as follows. In Sec.~\ref{sec:LR} we briefly summarise the formalism for the computation of the asymmetric nuclear matter response to probes producing isospin flip. The results for several Skyrme functionals are presented and analysed in Sec.~\ref{sec:results}, where we discuss the effect of the spin-orbit and tensor terms on the response function, and also consider probes related to different isospin operators. In Sec.\ref{sec:err}, we discuss the impact of parameter uncertainties on the response functions. Finally, our main conclusions  are given in Sec.~\ref{sec:conclu}. 


\section{Linear response formalism}
\label{sec:LR}

The LR formalism for asymmetric nuclear matter has already been presented in previous articles~\cite{dav14b,report} but for no charge-exchange processes. Our aim here is to extend the calculation to probes represented by external fields of the form
\begin{equation}\label{eq:probe}
\sum_j {\rm e}^{i {\bf q} \cdot {\bf r}_j} \, \Theta_j^{(S)} \, \tau_j^{\pm} \, ,
\end{equation}
where $\Theta_j^{(S)}$ refers to the operator $\mathds{1}$ or $\sigma^z_j$, respectively for total spin $S=0, 1$, and $\tau^{\pm}_j$ are the usual isospin raising and lowering operators. The response functions for this kind of probes imply $(n,p)$ and $(p,n)$ charge exchange reactions, where a proton converts into a neutron and vice versa. More precisely, the operators $\tau^+$ and $\tau^-$ create particle-hole excitations of two different species, namely $pn^{-1}$ and $np^{-1}$, respectively. Both channels are such that the total isospin and its third component are equal to 1. In the following, we will focus on the operator $\tau^+$, which requires the ph interaction between $p n^{-1}$ states only. Whenever necessary, excitations created by $\tau^-$ may be simply obtained just by exchanging proton and neutron labels. 


\subsection{Residual interaction}

The residual interaction or ph interaction is currently defined as the second functional derivative of the energy functional~\cite{mig67,bro71} so that the matrix elements we are interested in can be obtained as:
\begin{equation}
\langle p n^{-1} | V_{ph} | p n^{-1} \rangle = \frac{\delta^2 \langle E \rangle}{\delta \rho_{np} \delta \rho_{pn}} \, ,
\end{equation}
where $\rho_{np}$ and $\rho_{pn}$ represents off-diagonal matrix elements of the one-body density. Because the Skyrme functional~\cite{per04} does not contain explicit charge exchange terms~\cite{bka18}, there are no rearrangement  contributions, so that the ph interaction coincides with the particle-particle interaction $\langle p n | V | n p \rangle$. Note that only the Pauli exchange term contributes to that interaction. 

Hereafter, the different ph channels will be labelled $(S,M)$ where $S$ stands for the total spin and $M$ its projection. Moreover, we will follow the standard notation~\cite{gar92,report} and indicate a general matrix element as $V_{ph}^{(SM; S'M')}({\bf k}_1, {\bf k}_2, {\bf q})$, where ${\bf k}_1, {\bf k}_2$ are the hole momenta, ${\bf q}$ is the transferred momentum and we omit the proton and neutron indices in order to simplify the notation.  For the general Skyrme functional defined in Ref.~\cite{per04}, the ph interaction can be written as
\begin{eqnarray}\label{eq:res}
V_{ph}^{(SM; S'M')}({\bf k}_1, {\bf k}_2, {\bf q}) &=& \delta(S,S') \delta(M,M') \left[ W_{1}^{(S,1)}(q) 
      +  W_{2}^{(S,1)} \, ( {\bf k}_{1} - {\bf k}_{2})^{2} \right] \nonumber\\ 
& + & 8 C_{1}^{\nabla s}\delta_{SS'}\delta_{S1}\delta_{MM'}\delta_{M0}q^{2}\nonumber\\
& + & 4C_{1}^{F} \left\{(-)^{M}(k_{12})_{-M}(k_{12})_{M'}\delta_{SS'}\delta_{S1}-\frac{1}{2}\delta_{SS'}\delta_{S1}\delta_{M0}\delta_{M'0}q^{2}\right\}\nonumber\\
& + & 4qC_{1}^{\nabla J} \left( \delta_{S'0}\delta_{S1}M(k_{12})_{-M}+\delta_{S'1}\delta_{S0}M'(k_{12})_{M'}\right) \, ,
\end{eqnarray}
where the parameters $W_i^{(S,T=1)}$ are defined as
\begin{eqnarray}
\frac{1}{4}W_1^{(0,1)}&=&2C_1^{\rho,0}+2C_1^{\rho,\gamma}\rho_0^{\gamma}-\left[ 2C^{\Delta\rho}_1+\frac{1}{2}C_1^{\tau}\right]q^2\;,\\
\frac{1}{4}W_1^{(1,1)}&=&2C_1^{s,0}+2C_1^{s,\gamma}\rho_0^{\gamma}-\left[ 2C^{\Delta s}_1+\frac{1}{2}C_1^{T}\right]q^2\;,\\
\frac{1}{4}W_2^{(0,1)}&=&C_1^{\tau}\;,\\
\frac{1}{4}W_2^{(1,1)}&=&C_1^T \, .
\end{eqnarray}

The constants $C_{1}^{X}$, with $X=\Delta s, F, \dots$, are the coupling constants of the density functional. If the functional has been obtained from an effective interaction, these coupling constants can also be expressed in terms of the interaction parameters as shown in Ref.~\cite{report}. Finally, it is worth noticing that  the charge-exchange process is not coupled to the $nn^{-1}$ and $pp^{-1}$ excitations, contrarily to the non-isospin flip processes~\cite{dav14b}.


\subsection{The nuclear strength function}

The calculation of the response function requires the prior knowledge of the RPA ph propagator, which itself satisfies the following Bethe-Salpeter (BS) equation
\begin{eqnarray}
G^{(pn^{-1},SM)}_{RPA}(\mathbf{k}_{1},q,\omega)&=&G^{(pn^{-1})}_{HF}(\mathbf{k}_{1},q,\omega)\nonumber\\
&+&G^{(pn^{-1})}_{HF}(\mathbf{k}_{1},q,\omega) \sum_{S'M'} \int \frac{d^{3}\mathbf{k}_{2}}{(2 \pi)^3} V_{ph}^{(SM; S'M')}({\bf k}_1, {\bf k}_2, {\bf q}) G^{(pn^{-1},S'M')}_{RPA}(\mathbf{k}_{1},q,\omega) \;.
\end{eqnarray}
The Hartree-Fock (HF) propagator is independent of $(S,M)$. At variance with the non-isospin flip case~\cite{dav14b}, it has now the form
\begin{eqnarray}\label{ghf}
G^{(pn^{-1})}_{HF}(\mathbf{k},q,\omega)=\frac{n_n(\mathbf{k})-n_p(\mathbf{k+q})}{\omega-\left[\varepsilon_p(\mathbf{k}+\mathbf{q})-\varepsilon_n({\bf k}) \right]+i\eta}\;,
\end{eqnarray}
\noindent where $\varepsilon_{\tau=n,p}$ are the HF single particle energies
\begin{equation}
\varepsilon_{\tau}(k)=\frac{k^2}{2m^*_{\tau}}+U_{\tau}\;.
\end{equation}
\noindent As usual, $m^*_{\tau}$ stands for the effective mass and $U_{\tau}$ is the single particle potential~\cite{her97}, excluding the $k^2$-dependence which is absorbed into the effective mass. . The occupation number $n_{\tau}({\bf k})$ is either a step function $\theta(k_F^{(\tau)}-k)$ at zero temperature or a Fermi-Dirac distribution
\begin{equation}
n_{\tau}(\mathbf{k})=\left[ 1 + \exp^{(\varepsilon_{\tau}(k)-\mu_{\tau})/T}\right]^{-1} \, ,
\end{equation}
at finite temperature.

Once the BS equation is solved, we calculate the response function of the system as
\begin{eqnarray}
\chi^{(pn^{-1},SM)}(q,\omega)=2 \int \frac{d^3 {\bf k}}{(2 \pi)^3} G^{(pn^{-1},SM)}_{RPA}(\mathbf{k},q,\omega) \, .
\end{eqnarray}
Finally, the strength function is
\begin{equation}
S^{(pn^{-1},SM)}(q,\omega)=-\frac{1}{\pi}\frac{{\rm Im} \chi^{(pn^{-1},SM)}(q,\omega)}{1-e^{-\left(\tilde{\omega}-\mu_p+\mu_n\right)/T}}\;,
\end{equation}
 where $\tilde{\omega}=\omega-(U_p-U_n)$. We observe that using the detailed balance theorem we can relate the $pn^{-1}$ to $np^{-1}$ strength functions as
\begin{eqnarray}
S^{(pn^{-1},SM)}(q,\omega)=e^{-\left( \tilde{\omega}-\mu_p+\mu_n\right)/T}S^{(np^{-1},SM)}(q,-\omega)\;.
\end{eqnarray}

The method used to solve the BS equation has been discussed in Refs.~\cite{gar92,dav14b}. Essentially, it implies a closed linear system for several momentum integrals of the RPA propagator.  This linear system can then be cast in a matrix form as
\begin{eqnarray}
A X =B \, ,
\end{eqnarray}
\noindent where $A$ is the interaction matrix containing the ph matrix elements as well as momentum integrals of the HF propagator, $X$ contains the unknown momentum integrals of RPA propagators, including the response function and $B$ contains only momentum integrals of HF propagators. These matrices are explicitly given in Appendix~\ref{app:matrix}.


\section {Results}
\label{sec:results}

We now come to the presentation and discussion of the response functions at zero and finite temperature using  different functionals and for various isospin asymmetries, defined by the parameter
\begin{equation}
Y=\frac{\rho_n-\rho_p}{\rho_0}\;,
\end{equation}
\noindent where $\rho_{n(p)}$ is the neutron (proton)  density and $\rho_0=\rho_n+\rho_p$. The PNM case corresponds to the value $Y=1$, while the isospin saturated symmetric nuclear matter (SNM) is obtained for $Y=0$. Asymmetric nuclear matter (ANM) corresponds to intermediate values of $Y$. Hereafter, we will consider two representative asymmetries, namely $Y=0.21$ which corresponds roughly to the asymmetry of $^{208}$Pb and $Y=0.5$ which is the typical value of nuclear matter in $\beta$-equilibrium within a neutron star.

We present results obtained with the following Skyrme functionals: SLy5~\cite{cha97}, T22, T44~\cite{les07}, and Skxta~\cite{bro06}. The former contains central and spin-orbit terms, the other three also include a tensor term. The first three functionals have been derived using the same Saclay-Lyon fitting protocol while the fourth one was obtained with a different protocol. For the reasons explained below, we also present results for the central part of the Skyrme SGII interaction~\cite{gia81}. 


\subsection{Results with central terms only}

The formalism of the  charge exchange linear response theory has been already presented and discussed in Ref.~\cite{her99}, but limited to the central part of the SGII. Because our algorithm to obtain the response function is more complex that the one of Ref.~\cite{her99}, we have decided to perform the same calculations to benchmark our results. In this way we have detected that the value of parameter $x_3$ used in~\cite{her99}  was erroneously divided by a factor of ten\footnote{Incidentally we have also detected a misprint in Eq.~(24) of \cite{her99}, the factor $(k_F^2(-\tau)/k_F^2(\tau))^2$ has to be replaced by $k_F^2(-\tau)/k_F^2(\tau)$.}. We have thus checked that both algorithms give exactly the same results, provided the same parameters are introduced.

\begin{figure}[!h]
\begin{center}
\includegraphics[width=0.5\textwidth,angle=-90]{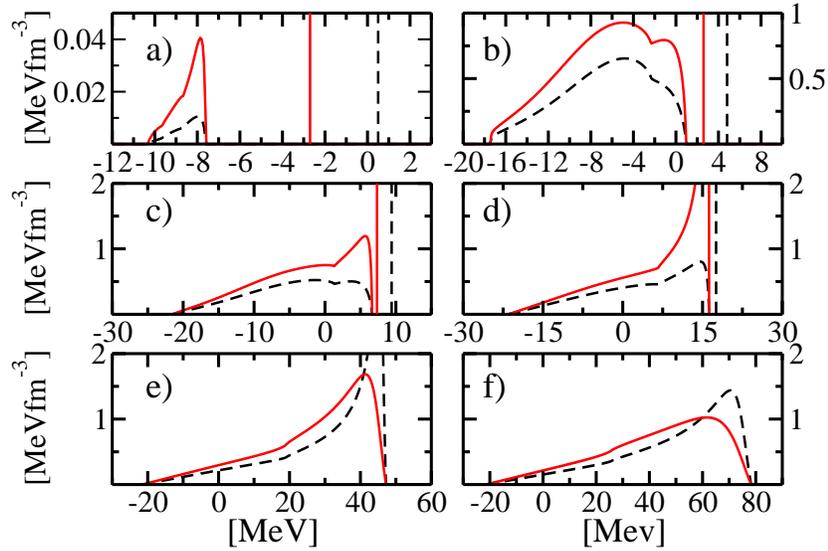}
\end{center}
\caption{(Color online). The imaginary part of the dynamical susceptibility  $-1000 \hbox{ Im } \chi^{{S,T}}(q,\omega)/\pi$ of asymmetric nuclear matter as a function of energy, for the Skyrme interaction SGII at saturation density of symmetric nuclear matter, $Y=0.2$  and zero temperature. Solid and dashed lines correspond to $S=1$ and 0 channels, respectively. The location of the collective states is indicated by vertical lines. The transferred momentum $q/k_F$ is (a) 0.01; (b) 0.09; (c) 0.14; (d) 0.22; (e) 0.45 and (f) 0.65.  }
\label{response:sgii}
\end{figure}

In Fig.~\ref{response:sgii}, we display the strength function obtained with the central part only of the SGII interaction at saturation density and $Y=0.2$ for spin channels $S=1$ (solid lines) and $S=0$ (dashed lines). The different panels refer to different values of transfer momentum $q/k_F$. A comparison with Fig.~3 of Ref.~\cite{her99} shows that although qualitatively in agreement, there are significant quantitative differences, in particular in the location of the collective states. 

As already stressed in Ref.~\cite{her99}, an important point is the presence at $T=0$ of collective oscillations at negative energies, for small momenta. This is related to the absence of Coulomb effects, which certainly modify the single particle spectrum for protons. A possible way to simulate Coulomb effects in infinite nuclear matter would be adding a repulsive shift to the proton mean field (see e.g. \cite{ose93}). In the present case, as the neutron Fermi energy lies higher than the proton one, $(n,p)$ transitions at low energy and momentum are possible in some cases. Such transitions do not exist in the case of ph excitations of the type $n p^{-1}$, as will be shown later on in Fig.~\ref{response:q116:r1:taup}.

The results displayed in Fig.~\ref{response:sgii} have been obtained by ignoring the spin-orbit term of interaction SGII. 
It was shown in Ref.~\cite{mar06} that the spin-orbit term couples the spin channels $S=0$ and 1, acting differently for channels with different $M$-component. However, the effect on the response function is rather small, even at momentum transfer relatively large as compared to the Fermi momentum. Indeed, by inspecting Eq.~\ref{w01}, we observe that the spin-orbit term contributes via a $q^4$-term and thus becomes negligible for small transferred momenta. This is not the case of the tensor which plays a role at both low and high transferred momenta~\cite{pas13b,dav14b}.  We thus consider now Skyrme functionals containing both spin-orbit and tensor terms.


\subsection{Results with the full interaction}\label{tzero}

\begin{figure}[!h]
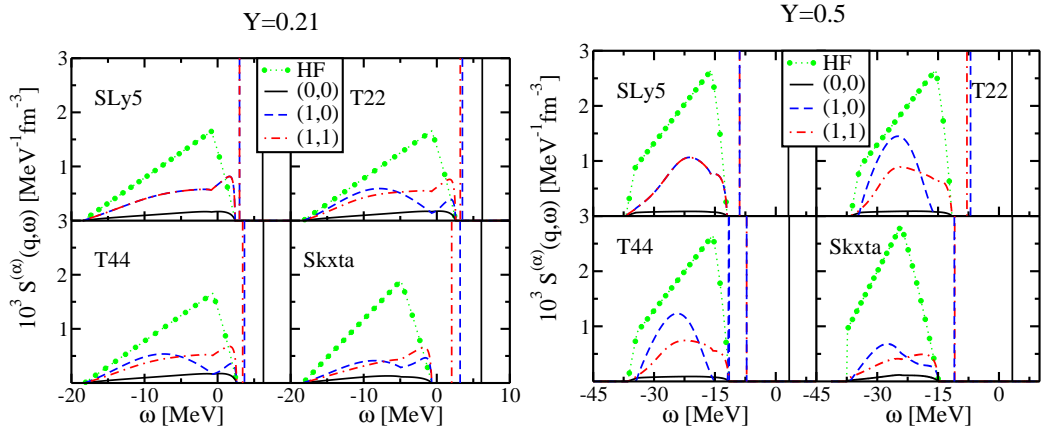

\begin{center}
\includegraphics[width=0.38\textwidth,angle=0]{fig2a.eps}
\includegraphics[width=0.38\textwidth,angle=0]{fig2b.eps}
\end{center}
\caption{(Color online). Strength function $S^{(\alpha)}(q,\omega)$ in asymmetric nuclear matter for the spin channels $\alpha=(S,M)$, using the four ph interactions considered here. Two different values of the asymmetry parameter are considered, $\rho/\rho_0=0.5$ and the transfer momentum is $q/k_F=0.2$. The Hartree-Fock strength function is also displayed (dotted line) to exhibit the effects of the ph interaction.}
\label{response:q0116}
\end{figure}

In Fig.~\ref{response:q0116}, are shown the strength functions obtained with the four considered Skyrme functionals, at  density $\rho/\rho_0=0.5$, transfer momentum $q/k_F=0.2$ and two values of the isospin asymmetries $Y$, for spin channels $(S,M)$. 
We observe that for SLy5 interaction, the only one with no tensor terms, the response functions in the channels (1,0) and (1,1) are fully degenerate. As discussed before, the spin-orbit term that should lift such an $M$-degeneracy is too small for the selected transferred momentum and thus no effect is visible. Since the tensor term in the residual interaction does not scale with $q^4$ but with $q^2$, it is active at lower transferred momentum. We thus notice that the other three Skyrme functionals (T22, T44, Skxta) break that degeneracy in the $S=1$ channel.

\begin{figure}[!h]
\begin{center}
\includegraphics[width=0.38\textwidth,angle=-90]{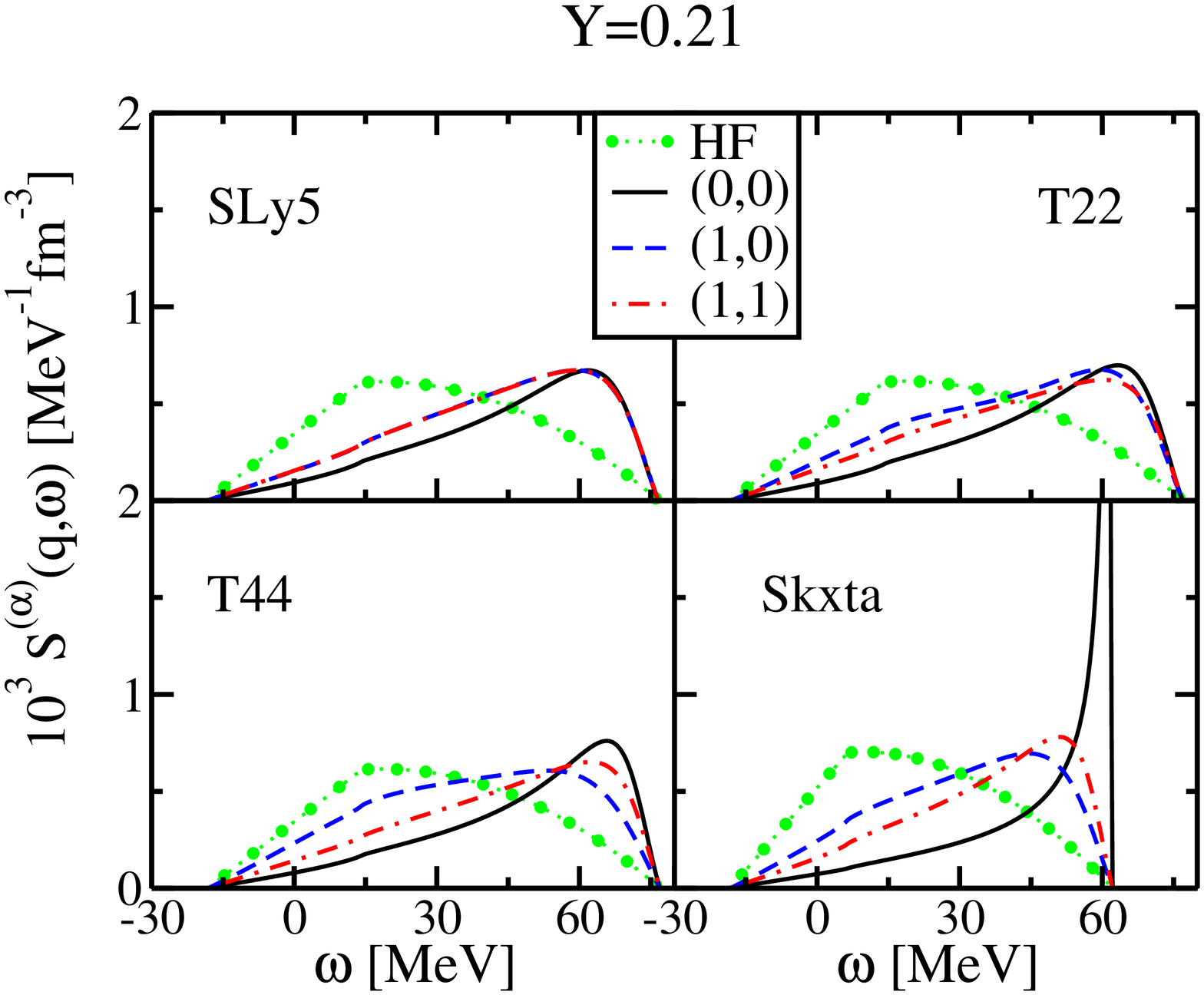}
\includegraphics[width=0.38\textwidth,angle=-90]{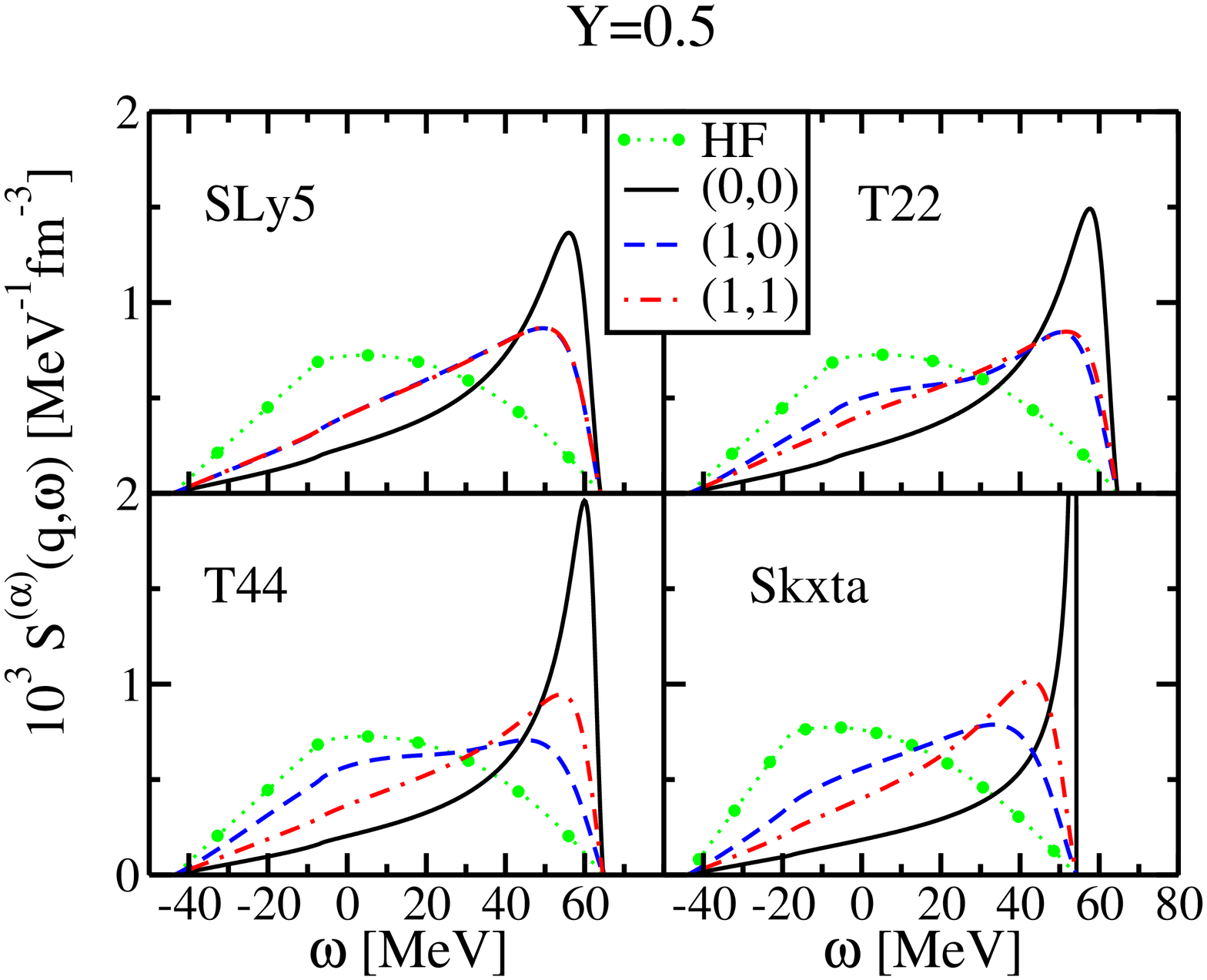}
\end{center}
\caption{(Color online). Same as Fig.\ref{response:q0116} but for $q/k_F=1$. }
\label{response:q116}
\end{figure}

Tensor effects can be small at small transferred momenta and small asymmetry (see Fig.~\ref{response:q0116}, left panel), but they are significantly enhanced with increasing asymmetry (see Fig.~\ref{response:q0116}, right panel) and/or transferred momentum (Fig.~\ref{response:q116}). This confirms the intrinsic importance of the tensor on the response functions also for charge-exchange processes. 
One could naively expect that the tensor interaction affects $S=1$ channels only. However, as the spin-orbit term couples both spin channels, it turns out that the tensor term actually acts also in the $S=0$ channel. The ph interaction should thus include both spin-orbit and tensor contributions. 

To show the global effect of the ph interaction itself, we have also displayed the HF strength function. We observe on  Fig.~\ref{response:q0116} that the interaction has indeed a strong effect: the strength function is shifted towards the high-energy region and a collective mode appears. For the same asymmetries, but for a higher value of the transferred momentum (see Fig.~\ref{response:q116}), the zero-sound mode is re-absorbed in the continuum part of the response. This is clearly seen in right panel at $Y=0.5$ where the strengths is the $S=0$ channel present a strong peak at the edge of the allowed region. If we now come to the tensor, we see that the effect is small for small transferred momenta ($q^2$ coupling) whatever the asymmetry is (see Fig.~\ref{response:q0116}) but becomes more pronounced at higher transferred momentum (see Fig.~\ref{response:q116}) where it increases the accumulation of strength at high energy.

\begin{figure}[!h]
\begin{center}
\includegraphics[width=0.38\textwidth,angle=-90]{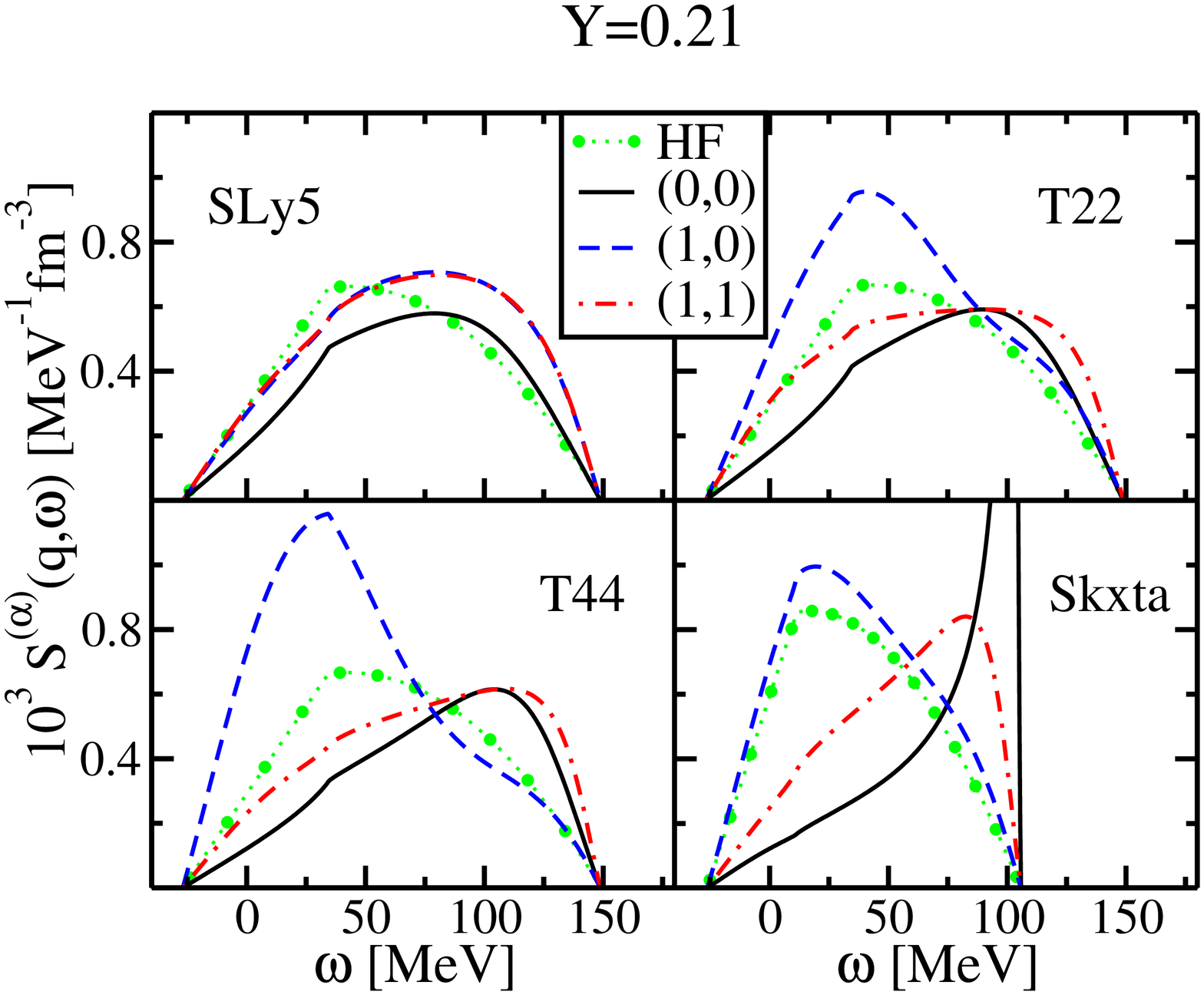}
\includegraphics[width=0.38\textwidth,angle=-90]{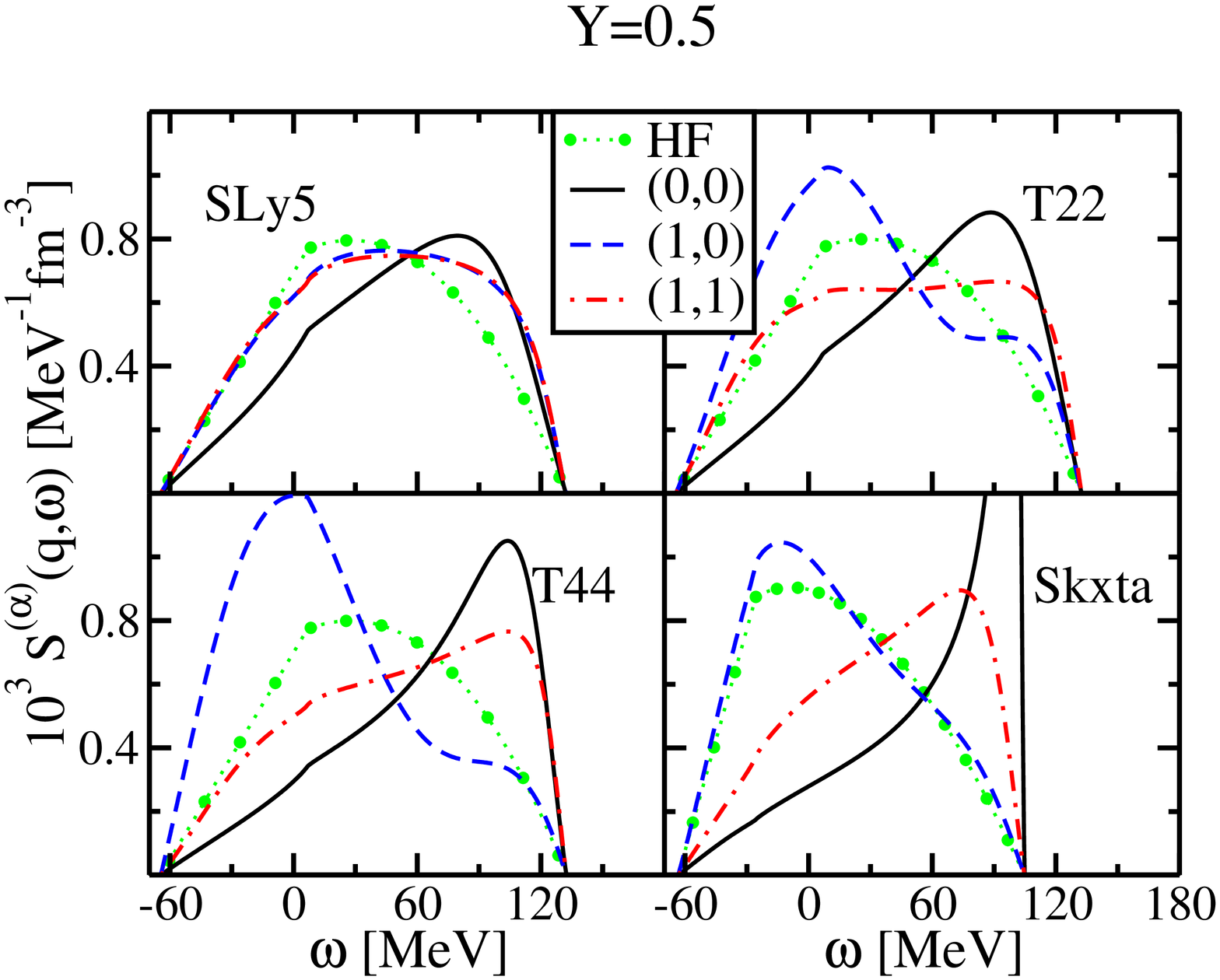}
\end{center}
\caption{(Color online). Same as Fig.\ref{response:q0116} but for $q/k_F=1$ and $\rho/\rho_0=1$. }
\label{response:q116:r1}
\end{figure}

The density is also an important parameter. To quantify its effect, we reported in Fig.~\ref{response:q116:r1} the strength functions at saturation density. In this case the effect of the tensor in the two spin-projection channels is opposite for all functionals: in the $M=0$ channel the tensor leads to a strong attraction and thus an accumulation of strength at low energy, while the opposite is true in the $M=1$ channel. As expected, such an effect is absent in SLy5 since it does not contain an explicit tensor term, and the two curves lie on top of each other.

\begin{figure}[!h]
\begin{center}
\includegraphics[width=0.38\textwidth,angle=-90]{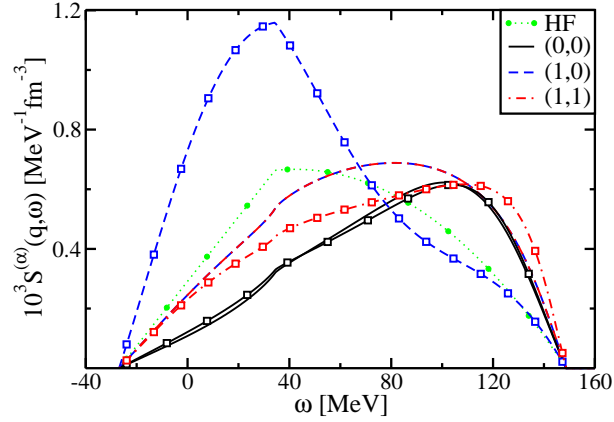}
\end{center}
\caption{(Color online). Response function  for T44 using  $q/k_F=1$ and $\rho/\rho_0=1$ and Y=0.21. The solid lines refer to the calculation without tensor  while open symbols refer to full interaction.}
\label{response:q116:r1:notens}
\end{figure}

One can also assess the role of the tensor contribution by comparing SLy5, T22 and T44 results in Fig.~\ref{response:q116:r1}. As discussed in Ref~\cite{les07} the $T_{ij}$ interactions have been fitted with very similar protocol as SLy5, but including an explicit tensor term. In order to get some insight of the role played by the tensor, we have thus repeated the calculations with T44 only,  turning off (on) the tensor coupling constants $C^F_1$ and $C^{\nabla s}_1$ The result is illustrated in Fig.~\ref{response:q116:r1:notens} for Y=0.21 at saturation density and transferred momentum  $q/k_F=1$.

We clearly observe that the tensor interaction acts in all channels and that its effect is quite remarkable especially in the (1,0) channel. In this case we notice that the calculations without tensor (solid line) exhibit an accumulation of strength at high energy, thus meaning a repulsive residual interaction.As previously discussed, in this case the (1,0) and (1,1) are essentially degenerate since the spin-orbit contribution is too small at low transferred momenta to provide a visible effect. 
When the tensor is taken into account the response function (dashed lines) accumulates at low energy thus meaning a strongly attractive interaction. As already discussed previously in Refs~\cite{dav09,pas12,pas12a,pas13b,dav14b} the tensor has a very important role in determining the excited states of a nuclear system. 


\subsection{Thermal effects}\label{tfinite}

\begin{figure}[!h]
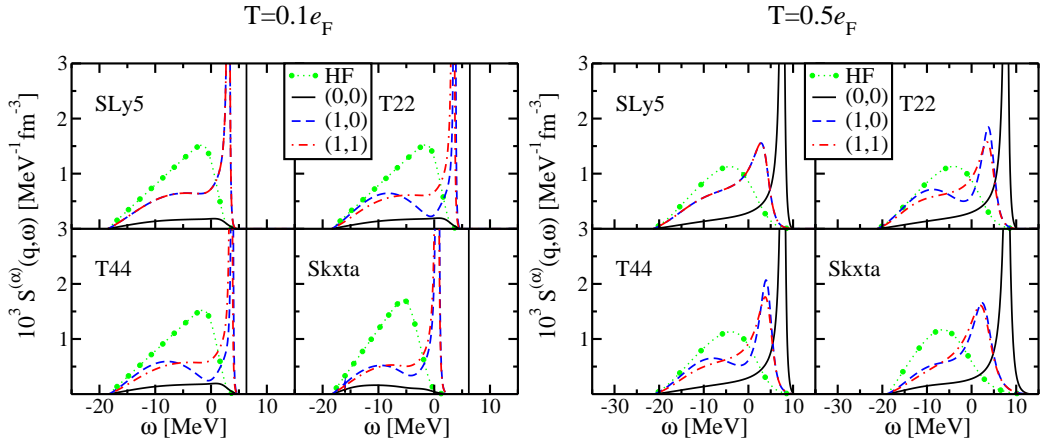

\begin{center}
\includegraphics[width=0.38\textwidth,angle=0]{fig6a.eps}
\includegraphics[width=0.38\textwidth,angle=0]{fig6b.eps}
\end{center}
\caption{(Color online). Same as Fig.\ref{response:q0116} at Y=0.21, but at $T=0.1e_F$ (left panel) and $T=0.5e_F$ (right panel). }
\label{response:q0116T}
\end{figure}

In Fig.~\ref{response:q0116T}, we illustrate the impact of temperature on the response function. For simplicity, we consider the same values of density and transferred momentum as in Fig.~\ref{response:q0116}, but now considering a finite temperature of  $T=0.1e_F$ and $T=0.5e_F$, where $e_F$ is the Fermi energy. We consider here only the case Y=0.21 since the other asymmetry leads to very similar results. As is well-known \cite{bra94,her96,her97,nav99,her99,report}, the effect of temperature is to wash out the structure of the response function and spread its strength. Moreover, a system at finite temperature can deexcite by giving some energy to the probe. These effects are clearly visible when comparing with Fig.~\ref{response:q0116}. The strength becomes broader, and in some cases the collective mode is absorbed into the continuum.

Going from $T=0.1e_F$ to $T=0.5e_F$  we observe that the limits of the strength function are increased and the peaks acquire a larger width. By further increasing the temperature, we may thus observe the complete disappearance of the peaks observed here and obtaining a smooth strength function over a large energy domain.
\subsection{Role of the isospin operator}\label{isospin}

Finally, we investigate the role of the isospin operator $\tau$ in the probe defined in Eq.~(\ref{eq:probe}).  In the case of asymmetric nuclear matter two  probes are possible: $\tau_z$ for non-isospin flip process and  $\tau^{\pm}$ for charge exchange excitations.
In the case of non-isospin flip probes, the relevant quantum numbers of each ph pair are the total spin ($S$), spin projection ($M$) and isospin ($T$). In the case of isospin-flip the ph-pairs are either $pn^{-1}$ (corresponding to the $\tau^+$ operator) or  $np^{-1}$ (for $\tau^-$). Also in this case the pairs are coupled to $S$ and $M$, the value of $T$ being equal to 1.

\begin{figure}[!h]
\begin{center}
\includegraphics[width=0.38\textwidth,angle=-90]{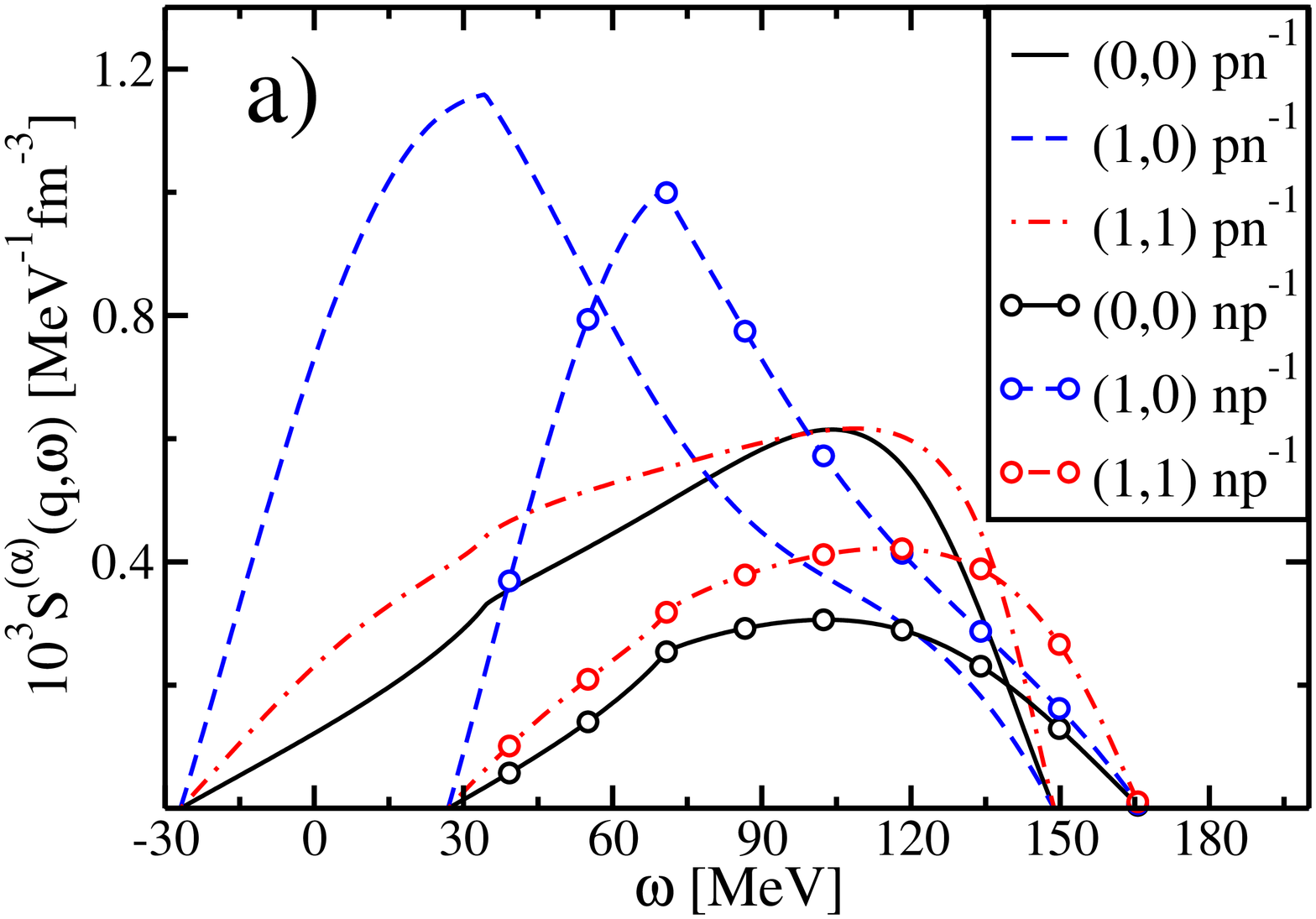}
\includegraphics[width=0.38\textwidth,angle=-90]{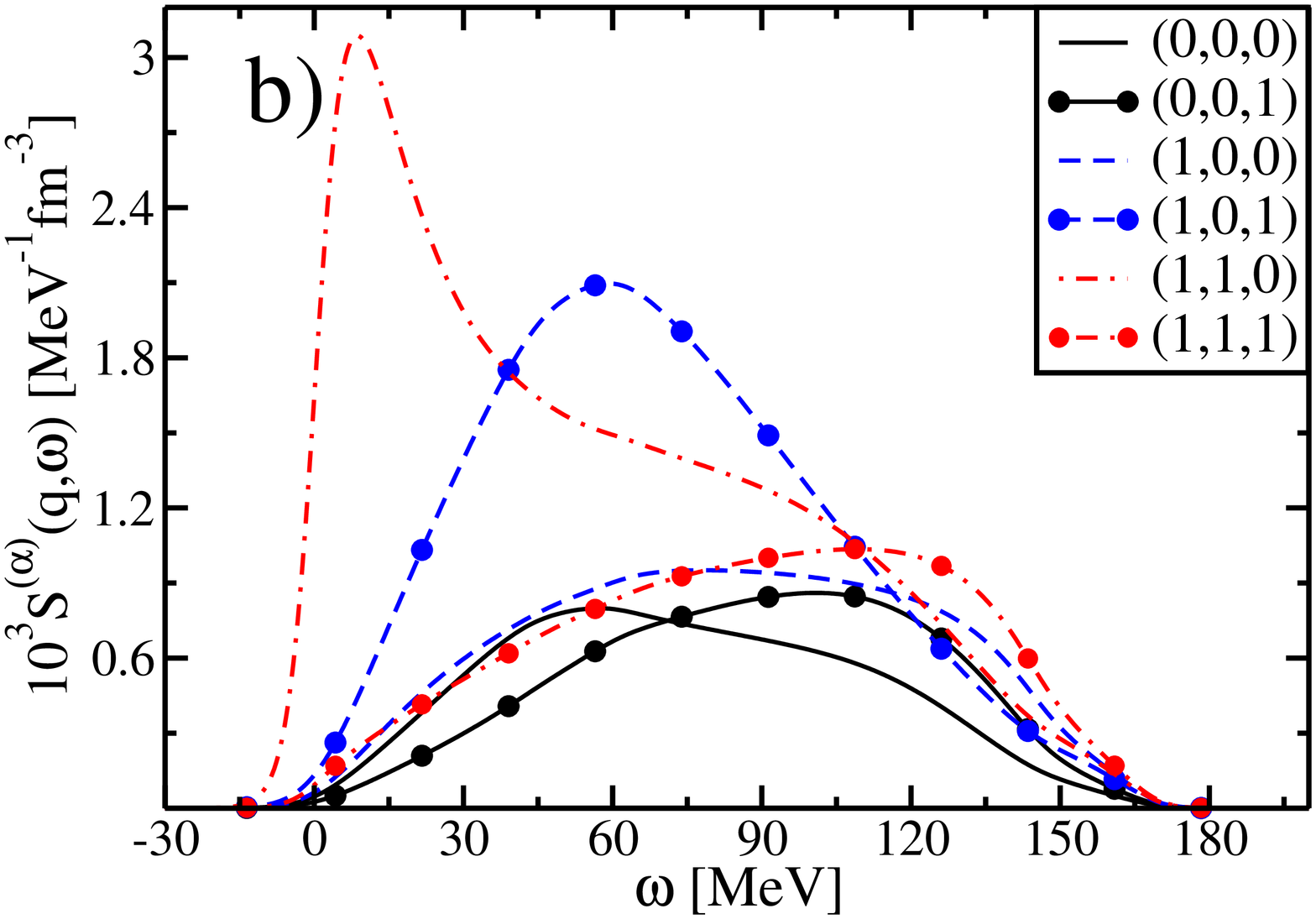}
\end{center}
\caption{(Color online). Response function  for T44 using  $q/k_F=1$ and $\rho/\rho_0=1$ and Y=0.21. Panel a) shows the charge-exchange cases, namely with operators $\tau^+$ (solid lines) and $\tau^-$ (open symbols), for the different $(S,M)$ channels. Panel b) shows the strength functions for the case with no charge exchange, for the different $(S,M,T)$ channels.}
\label{response:q116:r1:taup}
\end{figure}

In  Fig.~\ref{response:q116:r1:taup}, we illustrate the difference between the two probes for the T44 functional and asymmetry $Y=0.21$. We also fix the density of the system to  $\rho/\rho_0=1$ and  the transferred momentum to $q/k_F=1$.
In panel (a) of Fig.~\ref{response:q116:r1:taup}, we report the isospin-flip case for the two operators $\tau^+$ (solid lines) and $\tau^-$ (dashed lines). We observe that the domain of energy where the response function exists is quite different in the case $np^{-1}$ and $pn^{-1}$. This shift in the energy domain is due to the difference of chemical potential for the two species. Given the current asymmetry, the result simply shows that is more favourable to a  $pn^{-1}$ pair due to the large neutron excess of the system than the opposite process.

For completeness, we compare in panel (b) of Fig.~\ref{response:q116:r1:taup} the strength functions for the operators $\tau^+$ (solid lines) and $\tau^z$ (dashed lines), under the same conditions as panel (a). Apart from the fact that there are now more channels, due to the degeneracy breaking of $M$, the main difference is that the strength is considerably reduced in the charge exchange channels. 
We may thus expect this to have an impact on astrophysical observables such as NMFP in dense stellar matter~\cite{iwa82}.

\section{Error analysis}\label{sec:err}

In this section, we provide the first quantitative analysis of statistical uncertainties on the nuclear response function.
Following Refs~\cite{dob14,roc15}, the error $V_y$ on a given observable $y(\mathbf{x},\mathbf{a})$ depending on some independent variable $\mathbf{x}$ and a set of parameters $\mathbf{a}$ is obtained as

\begin{eqnarray}\label{eq:err}
V_y(\mathbf{x})=\sum_{ij}\frac{\partial y(\mathbf{x},\mathbf{a})}{\partial a_i} \mathcal{C}_{ij}\frac{\partial y(\mathbf{x},\mathbf{a})}{\partial a_j} 
\end{eqnarray}

\noindent where $\mathcal{C}$ is the covariance matrix~\cite{bar93}. The partial derivative respect to parameter space in Eq.\ref{eq:err} are done using a finite-difference method as discussed in Ref.~\cite{hav17}.
To perform such calculations a critical ingredient is thus represented by $\mathcal{C}$. Unfortunately very few Skyrme functionals provide published values for the covariance matrix. In the following we will restrict to the UNEDF0 ~\cite{kor10} and UNEDF1 ~\cite{kor12} functionals, since all relevant statistical informations required to perform error propagations are available.

The only limitation of this functionals is that they have been explicitly fitted taking into account only time-even terms of the functional~\cite{per04}. This means that essentially the $S=1$ channel of the response function is not determined, we thus consider the response function of the system only in the S=0 channel.

\begin{figure}[!h]
\begin{center}
\includegraphics[width=0.45\textwidth,angle=0]{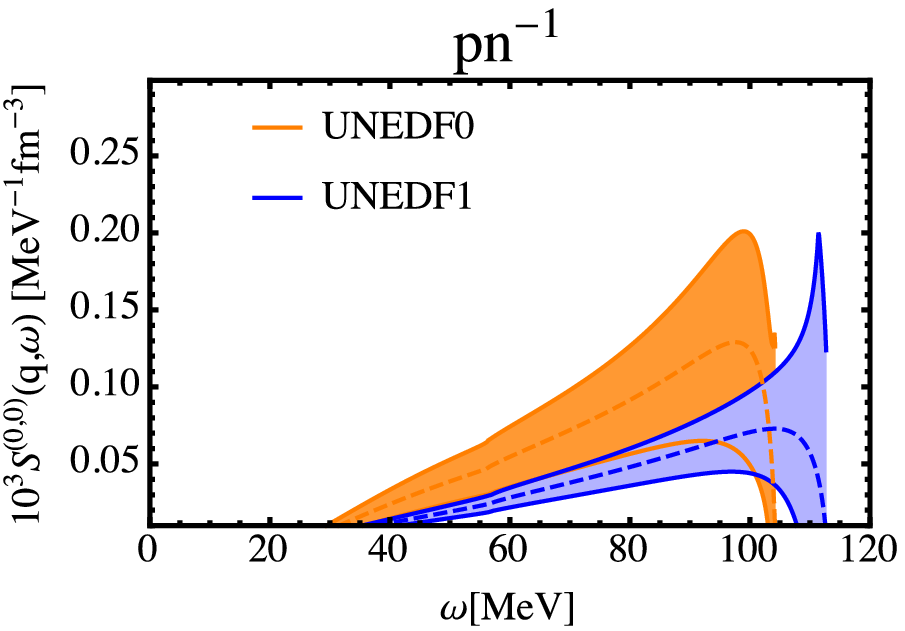}
\includegraphics[width=0.45\textwidth,angle=0]{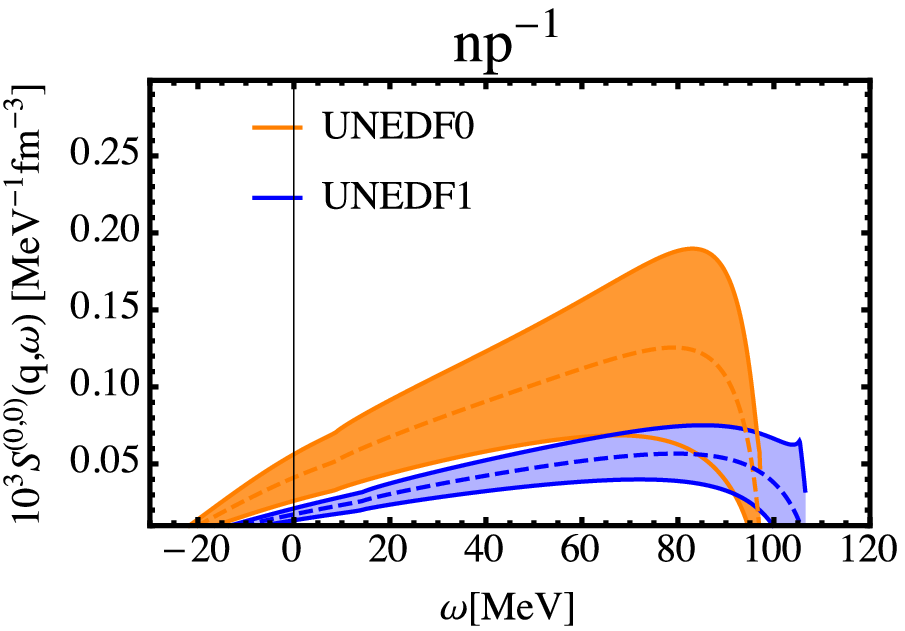}
\end{center}
\caption{(Colour online). Response function for UNEDF0 and UNEDF1 in the S=0 channel (dashed line). The colour bands represent the 1$\sigma$ error due to the statistical uncertainties in the parameters of the functional. See text for details}
\label{error}
\end{figure}

In Fig.\ref{error}, we show the response function $S^{(0,0)}(q,\omega)$ (dashed line) obtained with UNEDF0 and UNEDF1 for the charge exchange operator $\tau^-$ (left panel) and  $\tau^+$ (right panel).
By performing the full error propagation as defined in Eq.\ref{eq:err}, we have obtained the coloured bands appearing in the figure.
The band has been drawn to represent one standard deviation.

We observe that the main features of the response function are not much impacted by the statistical uncertainties, \emph{i.e} the attractive/repulsive structure of the response function is not affected by error. The latter have a strong impact on the actual height of the peaks.
This analysis is relevant for any model taking the response functions as input. For example, the calculations on NMFP may be strongly impacted by these error bars, but further investigations are required.


\section{Conclusions}
\label{sec:conclu}

In this paper we have generalised the LR formalism presented in Ref.~\cite{her99} so that Skyrme spin-orbit and tensor terms can be included. Moreover, the response functions were calculated for a general Skyrme functional ~\cite{per04}, thus providing more flexibility to study the behaviour of some particular coupling constants. We have investigated the evolution of the strength function as a function of the density of the system and the transferred momentum for some representative Skyrme functionals. We have observed that the presence of an explicit tensor term induces major effects on the strength function. 
 
Finally, we have generalised the formalism to the case of arbitrary isospin asymmetry and temperature so that the current methodology may be easily adopted to perform calculations of astrophysical interest as neutrino mean free path.
In a recent work~\cite{pas12a}, we have illustrated the role of the tensor on NMFP for the pure neutron matter case. From the results presented in the current article, we have shown that the tensor play a crucial role on the strength functions also in the case of isospin-flip probes. We thus may expect to observe a noticeable impact also in NMFP calculated in asymmetric matter. We leave this aspect for a future investigation.



\section*{Acknowledgements}
The work of JN was  supported by grant FIS2017-84038-C2-1P, Mineco (Spain). The work of AP was  funded by a grant from the UK Science and Technology Facilities Council (STFC): Consolidated Grant  ST/P003885/1. 


\begin{appendix}

\section{$\beta^{pn^{}}$ functions}\label{beta:functions}

To transform the Bethe-Salpeter equation into a closed linear system of algebraic equations, the propagators have to be integrated over the momentum with some weights. The HF propagator appears in the following integrals:
\begin{eqnarray}
\label{betafunct}
\beta_{i}^{pn}(q,\omega,T)&=&\int \frac{d^{3}k}{(2\pi)^{3}}G_{HF}^{pn^{-1}}(\mathbf{k},\mathbf{q},\omega,T)F_{i}(\mathbf{k},\mathbf{q})\\
F_{i}(\mathbf{k},\mathbf{q})&=&1,\frac{\mathbf{k}\cdot \mathbf{q}}{q^{2}},\frac{k^{2}}{q^{2}},\left[\frac{\mathbf{k}\cdot \mathbf{q}}{q^{2}}\right]^{2},\frac{(\mathbf{k}\cdot \mathbf{q})k^{2}}{q^{4}},\frac{k^{4}}{q^{4}},\left[\frac{\mathbf{k}\cdot \mathbf{q}}{q^{2}}\right]^{3},\left[\frac{\mathbf{k}\cdot \mathbf{q}}{q^{2}}\right]^{4},\frac{(\mathbf{k}\cdot \mathbf{q})^{2}k^{2}}{q^{6}} \, . \nonumber
\end{eqnarray}
\noindent The imaginary part of these integrals has been already presented in Ref.~\cite{report}. The real part is obtained numerically using the dispersion relation
\begin{eqnarray}
{\rm Re} \, \beta_{i}^{pn}(q,\omega,T) =-\frac{1}{\pi} \int_{-\infty}^{+\infty}d\omega' \frac{{\rm Im} \,  \beta_{i}^{pn}(q,\omega',T)}{\omega-\omega'} \, .
\end{eqnarray}


\section{Matrix elements}\label{app:matrix}

We give here the explicit expressions of the matrices required to determine the strength functions.


\subsection{Channel S=0}

The interaction matrix reads
\begin{eqnarray*}
A=\left( \begin{matrix}
1-\beta_{0}^{pn}\tilde{W}_{1}^{01}-q^{2}\beta_{2}^{pn}W_{2}^{01}&
-\beta_{0}^{pn^{-1}}W_{2}^{01}&
2q \beta_{1}^{pn^{-1}}W_{2}^{01}\\
-q^{2}\beta_{2}^{pn}\tilde{W}_{1}^{01}-q^{4}\beta_{5}^{pn}W_{2}^{01} &
1-q^{2}\beta_{2}^{pn}W_{2}^{01} &
2q^{3}\beta_{4}^{pn}W_{2}^{01} \\
-q\beta_{1}^{pn}\tilde{W}_{1}^{01}-q^{3}\beta_{4}^{pn}W_{2}^{01} &
-q\beta_{1}^{pn} W_{2}^{01}&
1 +2q^{2}\beta_{3}^{pn}W_{2}^{01}\\
\end{matrix}\right)
\end{eqnarray*}
where
\begin{eqnarray}\label{w01}
\tilde{W}^{01}_1=W^{01}_1+\frac{16q^4 \left(C^{\nabla J}_1\right)^2(\beta_2^{np}-\beta_3^{np})}{1+q^2(\beta_2^{np}-\beta_3^{np})\left( W_2^{11}-2C_1^F\right)}
\end{eqnarray}

The other two matrices read
\begin{eqnarray*}
X=\left( \begin{matrix}
\langle G^{pn^{-1},pn^{-1},00}_{RPA}\rangle\\
\langle k^{2} G^{pn^{-1},pn^{-1},00}_{RPA}\rangle\\ 
\sqrt{\frac{4\pi}{3}}\langle kY_{10}G^{pn^{-1},pn^{-1},00}_{RPA}\rangle\\ 
\end{matrix}\right)\;\;\;\;B=
\left( \begin{matrix}
\beta_{0}^{pn^{}}\\
q^{2}\beta_{2}^{pn^{}}\\
q\beta_{1}^{pn^{}}\\
\end{matrix}\right)
\end{eqnarray*}


\subsection{Channel S=1 M=$\pm$1}

\begin{scriptsize}
\begin{eqnarray*}
\mathbf{A}=\left( \begin{matrix}
1-\beta_{0}^{pn}\tilde{W}_{1}^{11}-q^{2}\beta_{2}^{pn}W_{2}^{11}&-\beta_{0}^{pn}W_{2}^{11}&
2q \beta_{1}^{pn}W_{2}^{11}&-2C^{F}_{1}\beta_{0}^{pn}\\
-2C^{F}_{1}q^{2}(\beta_{2}^{pn}-\beta_{3}^{pn})  & &-8q^{3}[C^{F}_{1}]^{2} \frac{\beta_{1}^{pn}(\beta_{2}^{pn}-\beta_{3}^{pn})-\beta_{0}^{pn}(\beta_{4}^{pn}-\beta_{6}^{pn})}{1+z^{pn,1}}  & \\ 
+8[C^{F}_{1}]^{2}q^{4}\beta_{1}^{pn}\frac{(\beta_{4}^{pn}-\beta_{6}^{pn})}{1+z^{pn,1}} & & & \\
 & & & \\
-q^{2}\beta_{2}^{pn}\tilde{W}_{1}^{11}-q^{4}\beta_{5}^{pn}W_{2}^{11} &
1-q^{2}\beta_{2}^{pn}W_{2}^{11}&
2q^{3} \beta_{4}^{pn}W_{2}^{11}+8q^{5}[C^{F}_{1}]^{2} \frac{\beta_{4}^{pn}\beta_{3}^{pn}-\beta_{2}^{pn}\beta_{6}^{pn}}{1+z^{pn,1}} &-2q^{2}C^{F}_{1}\beta_{2}^{pn}\\
-2C^{F}_{1}q^{4}(\beta_{5}^{pn}-\beta_{8}^{pn}) & & & \\
 & & & \\
 & & & \\
-q\beta_{1}^{pn}\tilde{W}_{1}^{11}-q^{3}\beta_{4}^{pn}W_{2}^{11} &
-q\beta_{1}^{pn}W_{2}^{11}&
1+2q^{2} \beta_{3}^{pn}W_{2}^{11}&-2qC^{F}_{1}\beta_{1}^{pn}\\
-2C^{F}_{1}q^{3}(\beta_{4}^{pn}-\beta_{6}^{pn}) & & -8q^{4}[C^{F}_{1}]^{2} \frac{\beta_{3}^{pn}(\beta_{2}^{pn}-\beta_{3}^{pn})-\beta_{1}^{pn}(\beta_{4}^{pn}-\beta_{6}^{pn})}{1+z^{pn,1}} & \\
+8[C^{F}_{1}]^{2}q^{5}\beta_{3}^{pn}\frac{(\beta_{4}^{pn}-\beta_{6}^{pn})}{1+z^{pn,1}}  & & & \\
& & & \\
-q^{2}(\beta_{2}^{pn}-\beta_{3}^{pn})\tilde{W}_{1}^{11}-q^{4}(\beta_{5}^{pn}-\beta_{8}^{pn})W_{2}^{11}&
-q^{2}(\beta_{2}^{pn}-\beta_{3}^{pn})W_{2}^{11}&
2q^{3} (\beta_{4}^{pn}-\beta_{6}^{pn})W_{2}^{11} &1-2q^{2}C^{F}_{1} (\beta_{2}^{pn}-\beta_{3}^{pn})\\
-2C^{F}_{1}q^{4}(\beta_{5}^{pn}-2\beta_{8}^{pn}+\beta_{7}^{pn}) & & & \\
 +8[C^{F}_{1}]^{2}q^{6}\frac{(\beta_{4}^{pn}-\beta_{6}^{pn})^{2}}{1+z^{pn,1}} & & & \\
 & & & \\
\end{matrix}\right)\\
\end{eqnarray*}
\end{scriptsize}
\noindent and 
\begin{eqnarray*}
\mathbf{X}=\left( \begin{matrix}
\langle G^{pn^{-1},pn^{-1},11}_{RPA}\rangle\\
\\
\langle k^{2} G^{pn^{-1},pn^{-1},11}_{RPA}\rangle\\ 
\\
\sqrt{\frac{4\pi}{3}}\langle kY_{10}G^{pn^{-1},pn^{-1},11}_{RPA}\rangle\\ 
\\
\frac{8\pi}{3}\langle k|Y_{11}|^{2}G^{pn^{-1},pn^{-1},11}_{RPA}\rangle\\
\end{matrix}\right)
; \,\,\,\mathbf{B}=
\left( \begin{matrix}
\beta_{0}^{pn}\\
\\
q^{2}\beta_{2}^{pn}\\
\\
q\beta_{1}^{pn}\\
\\
q^{2}(\beta_{2}^{pn}-\beta_{3}^{pn}) \\
\end{matrix}\right)
\end{eqnarray*}

\begin{eqnarray}
\tilde{W_{1}}^{11}&=&W_{1}^{11}+8\frac{q^{4}[C^{\nabla J}_{1}]^{2}}{1+z^{pn,0}}(\beta_{2}^{pn}-\beta_{3}^{pn})+q^{4}[C^{F}_{1}]^{2}\left[ 4(\beta_{5}^{pn}-\beta_{7}^{pn})-8\frac{W_{2}^{11}q^{2}(\beta_{4}^{pn}-\beta_{6}^{pn})^{2}}{1+z^{pn,1}}\right]
\end{eqnarray}

\begin{eqnarray}\label{zeta:chex}
z^{pn,S}=W_{2}^{S1}q^{2}(\beta_{2}^{pn}-\beta_{3}^{pn})
\end{eqnarray}


\subsection{Channel S=1 M=0}

\begin{small}

\begin{eqnarray*}
A=\left( \begin{matrix}
1-\beta_{0}^{pn}\tilde{W}_{1}^{10}-q^{2}\beta_{2}^{pn}W_{2}^{11}+q\beta_{1}^{pn}\alpha_{1}^{1,pn}-4C^{F}_{1}q^{2}\beta_{3}^{pn}& -\beta_{0}^{pn}W_{2}^{11}& 2q\beta_{1}^{pn}\alpha_{3}^{1,pn}+\beta_{0}^{pn}\alpha_{1}^{1,pn}&-4C^{F}_{1}\beta_{0}^{pn}\\
& & & \\
-q^{2}\beta_{2}^{pn}\tilde{W}_{1}^{10}-q^{4}\beta_{5}^{pn}W_{2}^{11}+q^{3}\beta_{4}^{pn}\alpha_{1}^{1,pn}-4C^{F}_{1}q^{4}\beta_{8}^{pn}&1 -q^{2}\beta_{2}^{pn}W_{2}^{11}& 2q^{3}\beta_{4}^{pn}\alpha_{3}^{1,pn}+q^{2}\beta_{2}^{pn}\alpha_{1}^{1,pn}&-4C^{F}_{1}q^{2}\beta_{2}^{pn}\\
& & & \\
-q\beta_{1}^{pn}\tilde{W}_{1}^{10}-q^{3}\beta_{4}^{pn}W_{2}^{11}+q^{2}\beta_{3}^{pn}\alpha_{1}^{1,pn}-4C^{F}_{1}q^{3}\beta_{6}^{pn}& -q\beta_{1}^{pn}W_{2}^{11}& 1+2q^{2}\beta_{3}^{pn}\alpha_{3}^{1,pn}+q^{}\beta_{1}^{pn}\alpha_{1}^{1,pn}&-4C^{F}_{1}q^{}\beta_{1}^{pn}\\
& & & \\
-q^{2}\beta_{3}^{pn}\tilde{W}_{1}^{10}-q^{4}\beta_{8}^{pn}W_{2}^{11}+q^{3}\beta_{6}^{pn}\alpha_{1}^{1,pn}-4C^{F}_{1}q^{4}\beta_{7}^{pn}& -q^{2}\beta_{3}^{pn}W_{2}^{11}& 2q^{3}\beta_{6}^{pn}\alpha_{3}^{1,pn}+q^{2}\beta_{3}^{pn}\alpha_{1}^{1,pn}&1-4C^{F}_{1}q^{2}\beta_{3}^{pn}\\
& & & \\
\end{matrix}\right)\\
\end{eqnarray*}
\end{small}

\noindent and 
\begin{eqnarray*}
\mathbf{X}=\left(\begin{matrix}
\langle G^{pn^{-1},pn^{-1},10}_{RPA} (q,\omega)\rangle\\
\\
\langle k^{2} G^{pn^{-1},pn^{-1},10}_{RPA} (q,\omega)\rangle\\
\\
\sqrt{\frac{4\pi}{3}}\langle k Y_{10}G^{pn^{-1},pn^{-1},10}_{RPA} (q,\omega)\rangle\\
\\
{\frac{4\pi}{3}} \langle k^{2}|Y_{10}|^{2}G^{pn^{-1},pn^{-1},10}_{RPA} (q,\omega)\rangle\\
\end{matrix}\right)
; \,\,\,\mathbf{B}=
\left( \begin{matrix}
\beta_{0}^{pn}\\
\\
q^{2}\beta_{2}^{pn}\\
\\
q\beta_{1}^{pn}\\
\\
q^{2}\beta_{3}^{pn} \\
\end{matrix}\right)
\end{eqnarray*}

\begin{eqnarray}
\alpha^{1,pn}_{1}&=&16 [C^{F}_{1}]^{2}\frac{q^{3}(\beta_{4}^{pn}-\beta_{6}^{pn})}{1+\bar{z}^{pn,1}}\\
\alpha^{1,pn}_{3}&=&W_{2}^{11}+4C^{F}_{1}-8[C^{F}_{1}]^{2}\frac{q^{2}(\beta_{2}^{pn}-\beta_{3}^{pn})}{1+\bar{z}^{pn,1}} 
\end{eqnarray}

\begin{eqnarray}
\tilde{W}_{1}^{10}=W_{1}^{11}+q^{2}(8C^{\nabla s}_{1}-2C^{F}_{1})+16[C^{F}_{1}]^{2}q^{3}\left[ q(\beta_{8}^{pn}-\beta_{7}^{pn})-q^{3}\frac{(\beta_{4}^{pn}-\beta_{6}^{pn})^{2}}{1+\bar{z}^{pn,1}}[W_{2}^{11}+6C^{F}_{1}]\right]
\end{eqnarray}

\begin{eqnarray}\label{zeta:chex}
\bar{z}^{pn,S}=(W_{2}^{S1}+6C^{F}_{1})q^{2}(\beta_{2}^{pn}-\beta_{3}^{pn})
\end{eqnarray}

\end{appendix}


\bibliography{biblio}

\begin{thebibliography}{46}
\expandafter\ifx\csname natexlab\endcsname\relax\def\natexlab#1{#1}\fi
\expandafter\ifx\csname bibnamefont\endcsname\relax
  \def\bibnamefont#1{#1}\fi
\expandafter\ifx\csname bibfnamefont\endcsname\relax
  \def\bibfnamefont#1{#1}\fi
\expandafter\ifx\csname citenamefont\endcsname\relax
  \def\citenamefont#1{#1}\fi
\expandafter\ifx\csname url\endcsname\relax
  \def\url#1{\texttt{#1}}\fi
\expandafter\ifx\csname urlprefix\endcsname\relax\def\urlprefix{URL }\fi
\providecommand{\bibinfo}[2]{#2}
\providecommand{\eprint}[2][]{\url{#2}}

\bibitem[{\citenamefont{Hirata et~al.}(1987)}]{hir87}
\bibinfo{author}{\bibfnamefont{K.}~\bibnamefont{Hirata}} \bibnamefont{et~al.},
  \bibinfo{journal}{Phys. Rev. Lett.} \textbf{\bibinfo{volume}{58}},
  \bibinfo{pages}{1490} (\bibinfo{year}{1987}).

\bibitem[{\citenamefont{Mezzacappa}(2005)}]{mez05}
\bibinfo{author}{\bibfnamefont{A.}~\bibnamefont{Mezzacappa}},
  \bibinfo{journal}{Annu. Rev. Nucl. Part. Sci.} \textbf{\bibinfo{volume}{55}},
  \bibinfo{pages}{467} (\bibinfo{year}{2005}).

\bibitem[{\citenamefont{Janka}(2012)}]{jan12}
\bibinfo{author}{\bibfnamefont{H.-T.} \bibnamefont{Janka}},
  \bibinfo{journal}{Annu. Rev. Nucl. Part. Sci.} \textbf{\bibinfo{volume}{62}},
  \bibinfo{pages}{407} (\bibinfo{year}{2012}).

\bibitem[{\citenamefont{Burrows}(2013)}]{bur13}
\bibinfo{author}{\bibfnamefont{A.}~\bibnamefont{Burrows}},
  \bibinfo{journal}{Rev. Mod. Phys.} \textbf{\bibinfo{volume}{85}},
  \bibinfo{pages}{245} (\bibinfo{year}{2013}).

\bibitem[{\citenamefont{Burrows and Lattimer}(1986)}]{bur86}
\bibinfo{author}{\bibfnamefont{A.}~\bibnamefont{Burrows}} \bibnamefont{and}
  \bibinfo{author}{\bibfnamefont{J.~M.} \bibnamefont{Lattimer}},
  \bibinfo{journal}{Astrophys. J.} \textbf{\bibinfo{volume}{307}},
  \bibinfo{pages}{178} (\bibinfo{year}{1986}).

\bibitem[{\citenamefont{Mayle et~al.}(1987)\citenamefont{Mayle, Wilson, and
  Schramm}}]{may87}
\bibinfo{author}{\bibfnamefont{R.}~\bibnamefont{Mayle}},
  \bibinfo{author}{\bibfnamefont{J.~R.} \bibnamefont{Wilson}},
  \bibnamefont{and} \bibinfo{author}{\bibfnamefont{D.~N.}
  \bibnamefont{Schramm}}, \bibinfo{journal}{Astrophys. J.}
  \textbf{\bibinfo{volume}{318}}, \bibinfo{pages}{288} (\bibinfo{year}{1987}).

\bibitem[{\citenamefont{Keil}(1994)}]{kei94}
\bibinfo{author}{\bibfnamefont{W.}~\bibnamefont{Keil}},
  \bibinfo{journal}{Progr. Part. Nucl. Phys.} \textbf{\bibinfo{volume}{32}},
  \bibinfo{pages}{105} (\bibinfo{year}{1994}).

\bibitem[{\citenamefont{Iwamoto and Pethick}(1982)}]{iwa82}
\bibinfo{author}{\bibfnamefont{N.}~\bibnamefont{Iwamoto}} \bibnamefont{and}
  \bibinfo{author}{\bibfnamefont{C.~J.} \bibnamefont{Pethick}},
  \bibinfo{journal}{Phys. Rev. D} \textbf{\bibinfo{volume}{25}},
  \bibinfo{pages}{313} (\bibinfo{year}{1982}).

\bibitem[{\citenamefont{Pastore
  et~al.}(2012{\natexlab{a}})\citenamefont{Pastore, Martini, Buridon, Davesne,
  Bennaceur, and Meyer}}]{pas12a}
\bibinfo{author}{\bibfnamefont{A.}~\bibnamefont{Pastore}},
  \bibinfo{author}{\bibfnamefont{M.}~\bibnamefont{Martini}},
  \bibinfo{author}{\bibfnamefont{V.}~\bibnamefont{Buridon}},
  \bibinfo{author}{\bibfnamefont{D.}~\bibnamefont{Davesne}},
  \bibinfo{author}{\bibfnamefont{K.}~\bibnamefont{Bennaceur}},
  \bibnamefont{and} \bibinfo{author}{\bibfnamefont{J.}~\bibnamefont{Meyer}},
  \bibinfo{journal}{Phys. Rev. C} \textbf{\bibinfo{volume}{86}},
  \bibinfo{pages}{044308} (\bibinfo{year}{2012}{\natexlab{a}}).

\bibitem[{\citenamefont{Pastore
  et~al.}(2014{\natexlab{a}})\citenamefont{Pastore, Martini, Davesne, Navarro,
  Goriely, and Chamel}}]{pas14bsk}
\bibinfo{author}{\bibfnamefont{A.}~\bibnamefont{Pastore}},
  \bibinfo{author}{\bibfnamefont{M.}~\bibnamefont{Martini}},
  \bibinfo{author}{\bibfnamefont{D.}~\bibnamefont{Davesne}},
  \bibinfo{author}{\bibfnamefont{J.}~\bibnamefont{Navarro}},
  \bibinfo{author}{\bibfnamefont{S.}~\bibnamefont{Goriely}}, \bibnamefont{and}
  \bibinfo{author}{\bibfnamefont{N.}~\bibnamefont{Chamel}},
  \bibinfo{journal}{Phys. Rev. C} \textbf{\bibinfo{volume}{90}},
  \bibinfo{pages}{025804} (\bibinfo{year}{2014}{\natexlab{a}}).

\bibitem[{\citenamefont{Pastore et~al.}(2015)\citenamefont{Pastore, Davesne,
  and Navarro}}]{report}
\bibinfo{author}{\bibfnamefont{A.}~\bibnamefont{Pastore}},
  \bibinfo{author}{\bibfnamefont{D.}~\bibnamefont{Davesne}}, \bibnamefont{and}
  \bibinfo{author}{\bibfnamefont{J.}~\bibnamefont{Navarro}},
  \bibinfo{journal}{Phys. Reports} \textbf{\bibinfo{volume}{563}},
  \bibinfo{pages}{1} (\bibinfo{year}{2015}).

\bibitem[{\citenamefont{Perli\ifmmode~\acute{n}\else \'{n}\fi{}ska
  et~al.}(2004)\citenamefont{Perli\ifmmode~\acute{n}\else \'{n}\fi{}ska,
  Rohozi\ifmmode~\acute{n}\else \'{n}\fi{}ski, Dobaczewski, and
  Nazarewicz}}]{per04}
\bibinfo{author}{\bibfnamefont{E.}~\bibnamefont{Perli\ifmmode~\acute{n}\else
  \'{n}\fi{}ska}}, \bibinfo{author}{\bibfnamefont{S.~G.}
  \bibnamefont{Rohozi\ifmmode~\acute{n}\else \'{n}\fi{}ski}},
  \bibinfo{author}{\bibfnamefont{J.}~\bibnamefont{Dobaczewski}},
  \bibnamefont{and}
  \bibinfo{author}{\bibfnamefont{W.}~\bibnamefont{Nazarewicz}},
  \bibinfo{journal}{Phys. Rev. C} \textbf{\bibinfo{volume}{69}},
  \bibinfo{pages}{014316} (\bibinfo{year}{2004}).

\bibitem[{\citenamefont{Lattimer and Prakash}(2000)}]{lattimer2000nuclear}
\bibinfo{author}{\bibfnamefont{J.~M.} \bibnamefont{Lattimer}} \bibnamefont{and}
  \bibinfo{author}{\bibfnamefont{M.}~\bibnamefont{Prakash}},
  \bibinfo{journal}{Physics Reports} \textbf{\bibinfo{volume}{333}},
  \bibinfo{pages}{121} (\bibinfo{year}{2000}).

\bibitem[{\citenamefont{Steiner et~al.}(2005)\citenamefont{Steiner, Prakash,
  Lattimer, and Ellis}}]{steiner2005isospin}
\bibinfo{author}{\bibfnamefont{A.~W.} \bibnamefont{Steiner}},
  \bibinfo{author}{\bibfnamefont{M.}~\bibnamefont{Prakash}},
  \bibinfo{author}{\bibfnamefont{J.~M.} \bibnamefont{Lattimer}},
  \bibnamefont{and} \bibinfo{author}{\bibfnamefont{P.~J.} \bibnamefont{Ellis}},
  \bibinfo{journal}{Phys. Reports} \textbf{\bibinfo{volume}{411}},
  \bibinfo{pages}{325} (\bibinfo{year}{2005}).

\bibitem[{\citenamefont{Lattimer and Prakash}(2007)}]{lattimer2007neutron}
\bibinfo{author}{\bibfnamefont{J.~M.} \bibnamefont{Lattimer}} \bibnamefont{and}
  \bibinfo{author}{\bibfnamefont{M.}~\bibnamefont{Prakash}},
  \bibinfo{journal}{Phys. Reports} \textbf{\bibinfo{volume}{442}},
  \bibinfo{pages}{109} (\bibinfo{year}{2007}).

\bibitem[{\citenamefont{Reddy et~al.}(1998)\citenamefont{Reddy, Prakash, and
  Lattimer}}]{red98}
\bibinfo{author}{\bibfnamefont{S.}~\bibnamefont{Reddy}},
  \bibinfo{author}{\bibfnamefont{M.}~\bibnamefont{Prakash}}, \bibnamefont{and}
  \bibinfo{author}{\bibfnamefont{J.~M.} \bibnamefont{Lattimer}},
  \bibinfo{journal}{Phys. Rev. D} \textbf{\bibinfo{volume}{58}},
  \bibinfo{pages}{013009} (\bibinfo{year}{1998}).

\bibitem[{\citenamefont{Shen et~al.}(2003)\citenamefont{Shen, Lombardo,
  Van~Giai, and Zuo}}]{she03}
\bibinfo{author}{\bibfnamefont{C.}~\bibnamefont{Shen}},
  \bibinfo{author}{\bibfnamefont{U.}~\bibnamefont{Lombardo}},
  \bibinfo{author}{\bibfnamefont{N.}~\bibnamefont{Van~Giai}}, \bibnamefont{and}
  \bibinfo{author}{\bibfnamefont{W.}~\bibnamefont{Zuo}},
  \bibinfo{journal}{Phys. Rev. C} \textbf{\bibinfo{volume}{68}},
  \bibinfo{pages}{055802} (\bibinfo{year}{2003}),
  \urlprefix\url{https://link.aps.org/doi/10.1103/PhysRevC.68.055802}.

\bibitem[{\citenamefont{Cowell and Pandharipande}(2004)}]{cow04}
\bibinfo{author}{\bibfnamefont{S.}~\bibnamefont{Cowell}} \bibnamefont{and}
  \bibinfo{author}{\bibfnamefont{V.~R.} \bibnamefont{Pandharipande}},
  \bibinfo{journal}{Phys. Rev. C} \textbf{\bibinfo{volume}{70}},
  \bibinfo{pages}{035801} (\bibinfo{year}{2004}),
  \urlprefix\url{https://link.aps.org/doi/10.1103/PhysRevC.70.035801}.

\bibitem[{\citenamefont{Roberts et~al.}(2012)\citenamefont{Roberts, Reddy, and
  Shen}}]{rob12}
\bibinfo{author}{\bibfnamefont{L.~F.} \bibnamefont{Roberts}},
  \bibinfo{author}{\bibfnamefont{S.}~\bibnamefont{Reddy}}, \bibnamefont{and}
  \bibinfo{author}{\bibfnamefont{G.}~\bibnamefont{Shen}},
  \bibinfo{journal}{Phys. Rev. C} \textbf{\bibinfo{volume}{86}},
  \bibinfo{pages}{065803} (\bibinfo{year}{2012}),
  \urlprefix\url{https://link.aps.org/doi/10.1103/PhysRevC.86.065803}.

\bibitem[{\citenamefont{Mart\'{\i}nez-Pinedo
  et~al.}(2012)\citenamefont{Mart\'{\i}nez-Pinedo, Fischer, Lohs, and
  Huther}}]{pin12}
\bibinfo{author}{\bibfnamefont{G.}~\bibnamefont{Mart\'{\i}nez-Pinedo}},
  \bibinfo{author}{\bibfnamefont{T.}~\bibnamefont{Fischer}},
  \bibinfo{author}{\bibfnamefont{A.}~\bibnamefont{Lohs}}, \bibnamefont{and}
  \bibinfo{author}{\bibfnamefont{L.}~\bibnamefont{Huther}},
  \bibinfo{journal}{Phys. Rev. Lett.} \textbf{\bibinfo{volume}{109}},
  \bibinfo{pages}{251104} (\bibinfo{year}{2012}),
  \urlprefix\url{https://link.aps.org/doi/10.1103/PhysRevLett.109.251104}.

\bibitem[{\citenamefont{Hern{\'a}ndez et~al.}(1999)\citenamefont{Hern{\'a}ndez,
  Navarro, and Polls}}]{her99}
\bibinfo{author}{\bibfnamefont{E.}~\bibnamefont{Hern{\'a}ndez}},
  \bibinfo{author}{\bibfnamefont{J.}~\bibnamefont{Navarro}}, \bibnamefont{and}
  \bibinfo{author}{\bibfnamefont{A.}~\bibnamefont{Polls}},
  \bibinfo{journal}{Nucl. Phys. A} \textbf{\bibinfo{volume}{658}},
  \bibinfo{pages}{327} (\bibinfo{year}{1999}).

\bibitem[{\citenamefont{Davesne et~al.}(2009)\citenamefont{Davesne, Martini,
  Bennaceur, and Meyer}}]{dav09}
\bibinfo{author}{\bibfnamefont{D.}~\bibnamefont{Davesne}},
  \bibinfo{author}{\bibfnamefont{M.}~\bibnamefont{Martini}},
  \bibinfo{author}{\bibfnamefont{K.}~\bibnamefont{Bennaceur}},
  \bibnamefont{and} \bibinfo{author}{\bibfnamefont{J.}~\bibnamefont{Meyer}},
  \bibinfo{journal}{Phys. Rev. C} \textbf{\bibinfo{volume}{80}},
  \bibinfo{pages}{024314} (\bibinfo{year}{2009}).

\bibitem[{\citenamefont{Pastore
  et~al.}(2012{\natexlab{b}})\citenamefont{Pastore, Davesne, Lallouet, Martini,
  Bennaceur, and Meyer}}]{pas12}
\bibinfo{author}{\bibfnamefont{A.}~\bibnamefont{Pastore}},
  \bibinfo{author}{\bibfnamefont{D.}~\bibnamefont{Davesne}},
  \bibinfo{author}{\bibfnamefont{Y.}~\bibnamefont{Lallouet}},
  \bibinfo{author}{\bibfnamefont{M.}~\bibnamefont{Martini}},
  \bibinfo{author}{\bibfnamefont{K.}~\bibnamefont{Bennaceur}},
  \bibnamefont{and} \bibinfo{author}{\bibfnamefont{J.}~\bibnamefont{Meyer}},
  \bibinfo{journal}{Phys. Rev. C} \textbf{\bibinfo{volume}{85}},
  \bibinfo{pages}{054317} (\bibinfo{year}{2012}{\natexlab{b}}).

\bibitem[{\citenamefont{Pastore
  et~al.}(2012{\natexlab{c}})\citenamefont{Pastore, Bennaceur, Davesne, and
  Meyer}}]{pas12b}
\bibinfo{author}{\bibfnamefont{A.}~\bibnamefont{Pastore}},
  \bibinfo{author}{\bibfnamefont{K.}~\bibnamefont{Bennaceur}},
  \bibinfo{author}{\bibfnamefont{D.}~\bibnamefont{Davesne}}, \bibnamefont{and}
  \bibinfo{author}{\bibfnamefont{J.}~\bibnamefont{Meyer}},
  \bibinfo{journal}{Int. J. Mod. Phys. E} \textbf{\bibinfo{volume}{21}},
  \bibinfo{pages}{1250040} (\bibinfo{year}{2012}{\natexlab{c}}).

\bibitem[{\citenamefont{Pastore
  et~al.}(2014{\natexlab{b}})\citenamefont{Pastore, Davesne, and
  Navarro}}]{pas13b}
\bibinfo{author}{\bibfnamefont{A.}~\bibnamefont{Pastore}},
  \bibinfo{author}{\bibfnamefont{D.}~\bibnamefont{Davesne}}, \bibnamefont{and}
  \bibinfo{author}{\bibfnamefont{J.}~\bibnamefont{Navarro}},
  \bibinfo{journal}{J. Phys. G} \textbf{\bibinfo{volume}{41}},
  \bibinfo{pages}{055103} (\bibinfo{year}{2014}{\natexlab{b}}).

\bibitem[{\citenamefont{Davesne et~al.}(2014)\citenamefont{Davesne, Pastore,
  and Navarro}}]{dav14b}
\bibinfo{author}{\bibfnamefont{D.}~\bibnamefont{Davesne}},
  \bibinfo{author}{\bibfnamefont{A.}~\bibnamefont{Pastore}}, \bibnamefont{and}
  \bibinfo{author}{\bibfnamefont{J.}~\bibnamefont{Navarro}},
  \bibinfo{journal}{Phys. Rev. C} \textbf{\bibinfo{volume}{89}},
  \bibinfo{pages}{044302} (\bibinfo{year}{2014}).

\bibitem[{\citenamefont{Migdal}(1967)}]{mig67}
\bibinfo{author}{\bibfnamefont{A.~B.} \bibnamefont{Migdal}},
  \emph{\bibinfo{title}{The Theory of Finite Fermi Systems}}
  (\bibinfo{publisher}{Wiley, New York}, \bibinfo{year}{1967}).

\bibitem[{\citenamefont{Brown}(1971)}]{bro71}
\bibinfo{author}{\bibfnamefont{G.}~\bibnamefont{Brown}}, \bibinfo{journal}{Rev.
  Mod. Phys.} \textbf{\bibinfo{volume}{18}}, \bibinfo{pages}{1}
  (\bibinfo{year}{1971}).

\bibitem[{\citenamefont{Garcia-Recio et~al.}(1992)\citenamefont{Garcia-Recio,
  Navarro, Nguyen, and Salcedo}}]{gar92}
\bibinfo{author}{\bibfnamefont{C.}~\bibnamefont{Garcia-Recio}},
  \bibinfo{author}{\bibfnamefont{J.}~\bibnamefont{Navarro}},
  \bibinfo{author}{\bibfnamefont{V.~G.} \bibnamefont{Nguyen}},
  \bibnamefont{and} \bibinfo{author}{\bibfnamefont{L.}~\bibnamefont{Salcedo}},
  \bibinfo{journal}{Ann. Phys. (NY)} \textbf{\bibinfo{volume}{214}},
  \bibinfo{pages}{293 } (\bibinfo{year}{1992}).

\bibitem[{\citenamefont{Baczyk et~al.}(2018)\citenamefont{Baczyk, Dobaczewski,
  Konieczka, Satu{\l}a, Nakatsukasa, and Sato}}]{bka18}
\bibinfo{author}{\bibfnamefont{P.}~\bibnamefont{Baczyk}},
  \bibinfo{author}{\bibfnamefont{J.}~\bibnamefont{Dobaczewski}},
  \bibinfo{author}{\bibfnamefont{M.}~\bibnamefont{Konieczka}},
  \bibinfo{author}{\bibfnamefont{W.}~\bibnamefont{Satu{\l}a}},
  \bibinfo{author}{\bibfnamefont{T.}~\bibnamefont{Nakatsukasa}},
  \bibnamefont{and} \bibinfo{author}{\bibfnamefont{K.}~\bibnamefont{Sato}},
  \bibinfo{journal}{Phys. Lett. B} \textbf{\bibinfo{volume}{778}},
  \bibinfo{pages}{178} (\bibinfo{year}{2018}).

\bibitem[{\citenamefont{Hern\'andez et~al.}(1997)\citenamefont{Hern\'andez,
  Navarro, and Polls}}]{her97}
\bibinfo{author}{\bibfnamefont{E.~S.} \bibnamefont{Hern\'andez}},
  \bibinfo{author}{\bibfnamefont{J.}~\bibnamefont{Navarro}}, \bibnamefont{and}
  \bibinfo{author}{\bibfnamefont{A.}~\bibnamefont{Polls}},
  \bibinfo{journal}{Nucl. Phys. A} \textbf{\bibinfo{volume}{627}},
  \bibinfo{pages}{460 } (\bibinfo{year}{1997}).

\bibitem[{\citenamefont{Chabanat et~al.}(1997)\citenamefont{Chabanat, Bonche,
  Haensel, Meyer, and Schaeffer}}]{cha97}
\bibinfo{author}{\bibfnamefont{E.}~\bibnamefont{Chabanat}},
  \bibinfo{author}{\bibfnamefont{P.}~\bibnamefont{Bonche}},
  \bibinfo{author}{\bibfnamefont{P.}~\bibnamefont{Haensel}},
  \bibinfo{author}{\bibfnamefont{J.}~\bibnamefont{Meyer}}, \bibnamefont{and}
  \bibinfo{author}{\bibfnamefont{R.}~\bibnamefont{Schaeffer}},
  \bibinfo{journal}{Nucl. Phys. A} \textbf{\bibinfo{volume}{627}},
  \bibinfo{pages}{710} (\bibinfo{year}{1997}).

\bibitem[{\citenamefont{Lesinski et~al.}(2007)\citenamefont{Lesinski, Bender,
  Bennaceur, Duguet, and Meyer}}]{les07}
\bibinfo{author}{\bibfnamefont{T.}~\bibnamefont{Lesinski}},
  \bibinfo{author}{\bibfnamefont{M.}~\bibnamefont{Bender}},
  \bibinfo{author}{\bibfnamefont{K.}~\bibnamefont{Bennaceur}},
  \bibinfo{author}{\bibfnamefont{T.}~\bibnamefont{Duguet}}, \bibnamefont{and}
  \bibinfo{author}{\bibfnamefont{J.}~\bibnamefont{Meyer}},
  \bibinfo{journal}{Phys. Rev. C} \textbf{\bibinfo{volume}{76}},
  \bibinfo{pages}{014312} (\bibinfo{year}{2007}).

\bibitem[{\citenamefont{Brown et~al.}(2006)\citenamefont{Brown, Duguet, Otsuka,
  Abe, and Suzuki}}]{bro06}
\bibinfo{author}{\bibfnamefont{B.}~\bibnamefont{Brown}},
  \bibinfo{author}{\bibfnamefont{T.}~\bibnamefont{Duguet}},
  \bibinfo{author}{\bibfnamefont{T.}~\bibnamefont{Otsuka}},
  \bibinfo{author}{\bibfnamefont{D.}~\bibnamefont{Abe}}, \bibnamefont{and}
  \bibinfo{author}{\bibfnamefont{T.}~\bibnamefont{Suzuki}},
  \bibinfo{journal}{Phys. Rev. C} \textbf{\bibinfo{volume}{74}},
  \bibinfo{pages}{061303} (\bibinfo{year}{2006}).

\bibitem[{\citenamefont{Giai and Sagawa}(1981)}]{gia81}
\bibinfo{author}{\bibfnamefont{N.~V.} \bibnamefont{Giai}} \bibnamefont{and}
  \bibinfo{author}{\bibfnamefont{H.}~\bibnamefont{Sagawa}},
  \bibinfo{journal}{Phys. Lett. B} \textbf{\bibinfo{volume}{106}},
  \bibinfo{pages}{379 } (\bibinfo{year}{1981}).

\bibitem[{\citenamefont{Oset et~al.}(1993)\citenamefont{Oset, Strottman, Toki,
  and Navarro}}]{ose93}
\bibinfo{author}{\bibfnamefont{E.}~\bibnamefont{Oset}},
  \bibinfo{author}{\bibfnamefont{D.}~\bibnamefont{Strottman}},
  \bibinfo{author}{\bibfnamefont{H.}~\bibnamefont{Toki}}, \bibnamefont{and}
  \bibinfo{author}{\bibfnamefont{J.}~\bibnamefont{Navarro}},
  \bibinfo{journal}{Phys. Rev. C} \textbf{\bibinfo{volume}{48}},
  \bibinfo{pages}{2395} (\bibinfo{year}{1993}).

\bibitem[{\citenamefont{Margueron et~al.}(2006)\citenamefont{Margueron,
  Van~Giai, and Navarro}}]{mar06}
\bibinfo{author}{\bibfnamefont{J.}~\bibnamefont{Margueron}},
  \bibinfo{author}{\bibfnamefont{N.}~\bibnamefont{Van~Giai}}, \bibnamefont{and}
  \bibinfo{author}{\bibfnamefont{J.}~\bibnamefont{Navarro}},
  \bibinfo{journal}{Phys. Rev. C} \textbf{\bibinfo{volume}{74}},
  \bibinfo{pages}{015805} (\bibinfo{year}{2006}).

\bibitem[{\citenamefont{Braghin and Vautherin}(1994)}]{bra94}
\bibinfo{author}{\bibfnamefont{F.}~\bibnamefont{Braghin}} \bibnamefont{and}
  \bibinfo{author}{\bibfnamefont{D.}~\bibnamefont{Vautherin}},
  \bibinfo{journal}{Phys. Lett. B} \textbf{\bibinfo{volume}{333}},
  \bibinfo{pages}{289} (\bibinfo{year}{1994}).

\bibitem[{\citenamefont{Hern\'andez et~al.}(1996)\citenamefont{Hern\'andez,
  Navarro, Polls, and Ventura}}]{her96}
\bibinfo{author}{\bibfnamefont{E.~S.} \bibnamefont{Hern\'andez}},
  \bibinfo{author}{\bibfnamefont{J.}~\bibnamefont{Navarro}},
  \bibinfo{author}{\bibfnamefont{A.}~\bibnamefont{Polls}}, \bibnamefont{and}
  \bibinfo{author}{\bibfnamefont{J.}~\bibnamefont{Ventura}},
  \bibinfo{journal}{Nucl. Phys. A} \textbf{\bibinfo{volume}{597}},
  \bibinfo{pages}{1 } (\bibinfo{year}{1996}).

\bibitem[{\citenamefont{Navarro et~al.}(1999)\citenamefont{Navarro,
  Hern\'andez, and Vautherin}}]{nav99}
\bibinfo{author}{\bibfnamefont{J.}~\bibnamefont{Navarro}},
  \bibinfo{author}{\bibfnamefont{E.~S.} \bibnamefont{Hern\'andez}},
  \bibnamefont{and}
  \bibinfo{author}{\bibfnamefont{D.}~\bibnamefont{Vautherin}},
  \bibinfo{journal}{Phys. Rev. C} \textbf{\bibinfo{volume}{60}},
  \bibinfo{pages}{045801} (\bibinfo{year}{1999}).

\bibitem[{\citenamefont{Dobaczewski et~al.}(2014)\citenamefont{Dobaczewski,
  Nazarewicz, and Reinhard}}]{dob14}
\bibinfo{author}{\bibfnamefont{J.}~\bibnamefont{Dobaczewski}},
  \bibinfo{author}{\bibfnamefont{W.}~\bibnamefont{Nazarewicz}},
  \bibnamefont{and} \bibinfo{author}{\bibfnamefont{P.}~\bibnamefont{Reinhard}},
  \bibinfo{journal}{Journal of Physics G: Nuclear and Particle Physics}
  \textbf{\bibinfo{volume}{41}}, \bibinfo{pages}{074001}
  (\bibinfo{year}{2014}).

\bibitem[{\citenamefont{Roca-Maza et~al.}(2015)\citenamefont{Roca-Maza, Paar,
  and Colo}}]{roc15}
\bibinfo{author}{\bibfnamefont{X.}~\bibnamefont{Roca-Maza}},
  \bibinfo{author}{\bibfnamefont{N.}~\bibnamefont{Paar}}, \bibnamefont{and}
  \bibinfo{author}{\bibfnamefont{G.}~\bibnamefont{Colo}},
  \bibinfo{journal}{Journal of Physics G: Nuclear and Particle Physics}
  \textbf{\bibinfo{volume}{42}}, \bibinfo{pages}{034033}
  (\bibinfo{year}{2015}).

\bibitem[{\citenamefont{Barlow}(1993)}]{bar93}
\bibinfo{author}{\bibfnamefont{R.~J.} \bibnamefont{Barlow}},
  \emph{\bibinfo{title}{Statistics: a guide to the use of statistical methods
  in the physical sciences}}, vol.~\bibinfo{volume}{29}
  (\bibinfo{publisher}{John Wiley \& Sons}, \bibinfo{year}{1993}).

\bibitem[{\citenamefont{Haverinen and Kortelainen}(2017)}]{hav17}
\bibinfo{author}{\bibfnamefont{T.}~\bibnamefont{Haverinen}} \bibnamefont{and}
  \bibinfo{author}{\bibfnamefont{M.}~\bibnamefont{Kortelainen}},
  \bibinfo{journal}{Journal of Physics G: Nuclear and Particle Physics}
  \textbf{\bibinfo{volume}{44}}, \bibinfo{pages}{044008}
  (\bibinfo{year}{2017}).

\bibitem[{\citenamefont{Kortelainen et~al.}(2010)\citenamefont{Kortelainen,
  Lesinski, Mor{\'e}, Nazarewicz, Sarich, Schunck, Stoitsov, and Wild}}]{kor10}
\bibinfo{author}{\bibfnamefont{M.}~\bibnamefont{Kortelainen}},
  \bibinfo{author}{\bibfnamefont{T.}~\bibnamefont{Lesinski}},
  \bibinfo{author}{\bibfnamefont{J.}~\bibnamefont{Mor{\'e}}},
  \bibinfo{author}{\bibfnamefont{W.}~\bibnamefont{Nazarewicz}},
  \bibinfo{author}{\bibfnamefont{J.}~\bibnamefont{Sarich}},
  \bibinfo{author}{\bibfnamefont{N.}~\bibnamefont{Schunck}},
  \bibinfo{author}{\bibfnamefont{M.}~\bibnamefont{Stoitsov}}, \bibnamefont{and}
  \bibinfo{author}{\bibfnamefont{S.}~\bibnamefont{Wild}},
  \bibinfo{journal}{Physical Review C} \textbf{\bibinfo{volume}{82}},
  \bibinfo{pages}{024313} (\bibinfo{year}{2010}).

\bibitem[{\citenamefont{Kortelainen et~al.}(2012)\citenamefont{Kortelainen,
  McDonnell, Nazarewicz, Reinhard, Sarich, Schunck, Stoitsov, and
  Wild}}]{kor12}
\bibinfo{author}{\bibfnamefont{M.}~\bibnamefont{Kortelainen}},
  \bibinfo{author}{\bibfnamefont{J.}~\bibnamefont{McDonnell}},
  \bibinfo{author}{\bibfnamefont{W.}~\bibnamefont{Nazarewicz}},
  \bibinfo{author}{\bibfnamefont{P.-G.} \bibnamefont{Reinhard}},
  \bibinfo{author}{\bibfnamefont{J.}~\bibnamefont{Sarich}},
  \bibinfo{author}{\bibfnamefont{N.}~\bibnamefont{Schunck}},
  \bibinfo{author}{\bibfnamefont{M.}~\bibnamefont{Stoitsov}}, \bibnamefont{and}
  \bibinfo{author}{\bibfnamefont{S.}~\bibnamefont{Wild}},
  \bibinfo{journal}{Physical Review C} \textbf{\bibinfo{volume}{85}},
  \bibinfo{pages}{024304} (\bibinfo{year}{2012}).

\end{thebibliography}

\end{document}